\documentclass[fleqn,usenatbib]{mnras}

\usepackage[T1]{fontenc}
\usepackage{ae,aecompl}

\usepackage{float}
\usepackage{graphicx}	
\usepackage{amsmath}	
\usepackage{amssymb}	
\graphicspath{{./}{figures/}}
\usepackage{xcolor}
\newcommand\hl[1]{%
  \bgroup
  \hskip0pt\color{black}
  #1%
  \egroup
}
\usepackage{adjustbox}

\title[Surfaces shape the climate of habitable exoplanets]{How surfaces shape the climate of habitable exoplanets}

\author[Madden \& Kaltenegger]{
Jack Madden,$^{1,2}$\thanks{E-mail: jmadden@astro.cornell.edu}
Lisa Kaltenegger,$^{1,2}$
\\
$^{1}$Astronomy and Space Sciences, Cornell University, Ithaca NY 14850, USA\\
$^{2}$Carl Sagan Institute, Ithaca NY 14850, USA\\
}

\date{Accepted 2020 February 5. Received 2020 January 27; in original form 2019 October 17.}

\pubyear{2019}

\begin{document}
\label{firstpage}
\pagerange{\pageref{firstpage}--\pageref{lastpage}}
\maketitle

\begin{abstract}
Large ground- and space-based telescopes will be able to observe Earth-like planets in the near future. 
We explore how different planetary surfaces can strongly influence the climate, atmospheric composition, and remotely detectable spectra of terrestrial rocky exoplanets in the habitable zone depending on the host star`s incident irradiation spectrum for a range of Sun-like host stars from F0V to K7V. 
We update a well-tested 1D climate-photochemistry model to explore the changes of a planetary environment for different surfaces for different host stars.
Our results show that using a wavelength-dependent surface albedo is critical for modeling potentially habitable rocky exoplanets.
\end{abstract}




\section{Introduction}
With dozens of Earth-sized planets already discovered, the next step in the search for life beyond our Solar System will be the characterization of the atmospheres of terrestrial planets and the search for signs of life on planets in the Habitable Zone (HZ). 

Gathering spectra of the atmospheres of potentially habitable exoplanets is one of the aims of several upcoming and proposed telescopes both on the ground and in space like the James Webb Space Telescope (JWST) and the Extremely Large telescopes (ELTs), such as the Giant Magellan Telescope (GMT), Thirty Meter Telescope (TMT), and the Extremely Large Telescope (ELT) and several missions concepts like Arial ~\citep{Tinetti2016}, Origins ~\citep{Battersby2018}, Habex ~\citep{Mennesson2016} and LUOVIR ~\citep{Luvoir2018}. Future ground-based ELTs and JWST are designed to be capable of obtaining first  measurements of the atmospheric composition of Earth-sized planets ~\cite[see e.g.][]{KalteneggerTraub2009,Kaltenegger2010,Stevenson2016,Barstow2016,Hedelt2013,Snellen2013,Rodler2014,Betremieux2014,Misra2014,Munoz2012}.

Surface reflection plays a critical role in a planet's climate due to the varying reflection of incoming starlight on the surface, depending on the surface composition. 1D models commonly used to simulate terrestrial exoplanet atmospheres such as, Chemclim ~\citep{Segura2010,Segura2005,Segura2003}, ATMOS ~\cite[see, e.g.][]{Arney2016}, and earlier versions of ExoPrime ~\citep{Rugheimer2018,Rugheimer2013} use a single, wavelength-independent albedo value for the planet, which has been calibrated to reproduce present-day Earth conditions for present-day Sun-like irradiance. 

Even though this average value for Earth's albedo is well calibrated to reproduce a similar climate to the wavelength-dependent albedo for present-day Earth orbiting the Sun, the particular average value will vary strongly if the incident stellar SED changes from a Sun-analog to a cool M star or hotter F star. For example, the cooling ice-albedo feedback is smaller for M stars than for G stars ~\citep{Shields2013,Abe2011} because ice reflects strongest in the visible shortwave region. In contrast, cool M stars emit most of their energy in the red part of the spectrum. The climate conditions driving the snowball-deglaciation loop in models show a dependence on the stellar type, which has an impact on long-term sustained surface habitability ~\citep{Abe2011,Shields2014,Abbot2018}. The relationship between the surface albedo and the stellar SED can lead to a substantial difference in the heating of a planet. We focus on F, G, and K-stars because M-stars pose a unique challenge for climate modeling due to the potential for planets to be tidally locked in the habitable zone and high stellar activity \citep{Airapetian2017,Johnstone2018}.

Several studies ~\citep{Kaltenegger2007,Cockell2009,Kaltenegger2010,Rugheimer2013,Rugheimer2015Spectra,Schwieterman2015,OMalleyJames2018,Rugheimer2018} have shown that surface albedo, as well as cloud coverage, is an essential factor for atmospheric as well as surface biosignature detection. 

However, no study has explored the feedback for a range of different surfaces on the climate, photochemistry, habitability, and observable spectra of planets in the HZ orbiting a wide range of stars. The modeled climate of a planet with a flat surface albedo only responds to differences in the total incident flux received whereas a planet with a non-flat surface albedo, responds to the wavelength dependence of that flux. Thus, a wavelength-dependent surface albedo is critical to assess the differing efficiencies of incoming stellar SED to heat or cool the planet. Depending on the surface, the effectiveness changes, and cannot be captured with one single value for the albedo at all wavelengths for stars with different SEDs.

Another critical issue that has not been explored is that the average surface albedo value commonly used encompasses the heating as well as cooling of clouds in addition to the reflectivity of a planet's surface. We address this by first exploring the contribution of clouds to the overall albedo for present-day Earth and separating that effect from the surface for present-day Earth. This separation is critical to be able to assess the influence of the surface environment on the planet's climate. Note that there is no self-consistent model that predicts the cloud feedback with different stellar types or stellar irradiance. Thus we keep the cloud component constant in our model comparison for different host stars, to isolate the effect of changing planetary surfaces. 

Our paper demonstrates the importance of including the wavelength-dependent feedback between a planet's surface and a planet's host star for Earth-like planets in the HZ. We focus on the change in a planet`s climate, its surface temperature as well as atmospheric species including biosignatures, which can indicate life on a planet: Ozone and oxygen in combination with a reducing gas like methane or N$_2$O ~\citep{Lovelock1965L,Lederberg1965,Lippincott1967}. Other atmospheric components we highlight are climate indicators like water and CO$_2$, which in addition to estimating the greenhouse gas concentration on an Earth-like planet, can also indicate whether the oxygen production can be explained abiotically  ~\cite[e.g.][]{DesMarais2002,Kaltenegger2017}. 
Section 2 describes our models, section 3 presents our results and section 4 a discussion. 

\section{Methods}
\subsection{Planetary Model}
A star's radiation shifts to longer wavelengths with cooler surface temperatures, which makes the light of a cooler star more efficient in heating an Earth-like planet with a mostly N$_2$-H$_2$O-CO$_2$ atmosphere ~\citep{Kasting1993}. This is partly due to the effectiveness of Rayleigh scattering, which decreases at longer wavelengths. A second effect is an increase in near-IR absorption by H$_2$O and CO$_2$ as the star's spectral peak shifts to these wavelengths. That means that the same integrated stellar flux that hits the top of a planet's atmosphere from a cool red star warms a planet more efficiently than from a hot blue star. Thus the stellar irradiance and the resulting orbital distance where a planet will show a similar surface temperature depends on the stellar host's SED. 

To establish the incident irradiation which produces similar surface temperatures for different host stars, we reduce the incident stellar flux at the moist greenhouse HZ limits for planet models with 1 Earth-mass ~\citep{Kopparapu2013} and fit it to the incident flux of present-day Earth for a G2V star to estimate the stellar irradiation for each stellar type.  The HZ is a concept that is used to guide remote observation strategies to characterize potentially habitable worlds. It is defined as the region around one or multiple stars in which liquid water could be stable on a rocky planet's surface ~\citep{Kasting1993,Kaltenegger2013,Kane2013,Kopparapu2013,Ramirez2016,Ramirez2017}, facilitating the remote detection of possible atmospheric biosignatures ~\cite[see e.g. review][]{Kaltenegger2017}.

This approach provides surface conditions similar to modern Earth of  $288K \pm 2\%$ across all stellar types for a wavelength-independent, fixed surface albedo (284 K for the K7V to 292K for the F0V host star). 

\subsection{Atmospheric Model}
For this study, we update exo-Prime to include a wavelength-dependent surface albedo. Exo-Prime ~\cite[see e.g.][]{Kaltenegger2010}, is a coupled 1D iterative radiative-convective atmosphere code with a line by line radiative transfer code, developed for rocky exoplanets. We update exo-Prime to include i) the updates in ATMOS ~\cite[see][]{Arney2016} in the climate and photochemical model as well as ii) a decoupled cloud and surface albedo and iii) wavelength-dependent albedo in all calculations, instead of a single average value.   

The code is based on iterations of a 1D climate  ~\citep{KastingAckerman1986,Pavlov2000,HaqqMisra2008}, and a 1D photochemistry model ~\citep{Pavlov2002,Segura2005,Segura2007}, which are run to convergence ~\cite[see details in][]{Segura2005}. Visible and near-IR shortwave fluxes are calculated with a two-stream approximation, including atmospheric gas scattering ~\citep{Toon1989}, and longwave fluxes in the IR region are calculated with a rapid radiative transfer model (RRTM). We use a geometrical model in which the average 1D global atmospheric model profile is generated using a plane-parallel atmosphere, treating the planet as a Lambertian sphere, and setting the stellar zenith angle to 60 degrees to represent the average incoming stellar flux on the dayside of the planet ~\cite[see also][]{SchindlerKasting2000}. A reverse-Euler method within the photochemistry code (originally developed by ~\cite{Kasting1985}) contains 220 reactions to solve for 55 chemical species. The radiative transfer code to model reflected planetary spectra in Exo-Prime was originally developed to study Earth spectra ~\citep{TraubStier1976} and later adapted for exoplanet use ~\citep{Kaltenegger2007,KalteneggerTraub2009}. We calculate light transmission at a resolution of $0.01 cm^{-1}$ from 0.4 to 2 microns providing a minimum resolving power of 100,000 at all wavelengths.

We divide the atmosphere into 100 layers for our model up to an altitude of at least 60 km, with smaller spacing towards the ground. We present the spectra at a resolution of 100 for clarity in the figures of this paper.

\subsection{Stellar Spectra}
\begin{figure}
\includegraphics[width=\columnwidth]{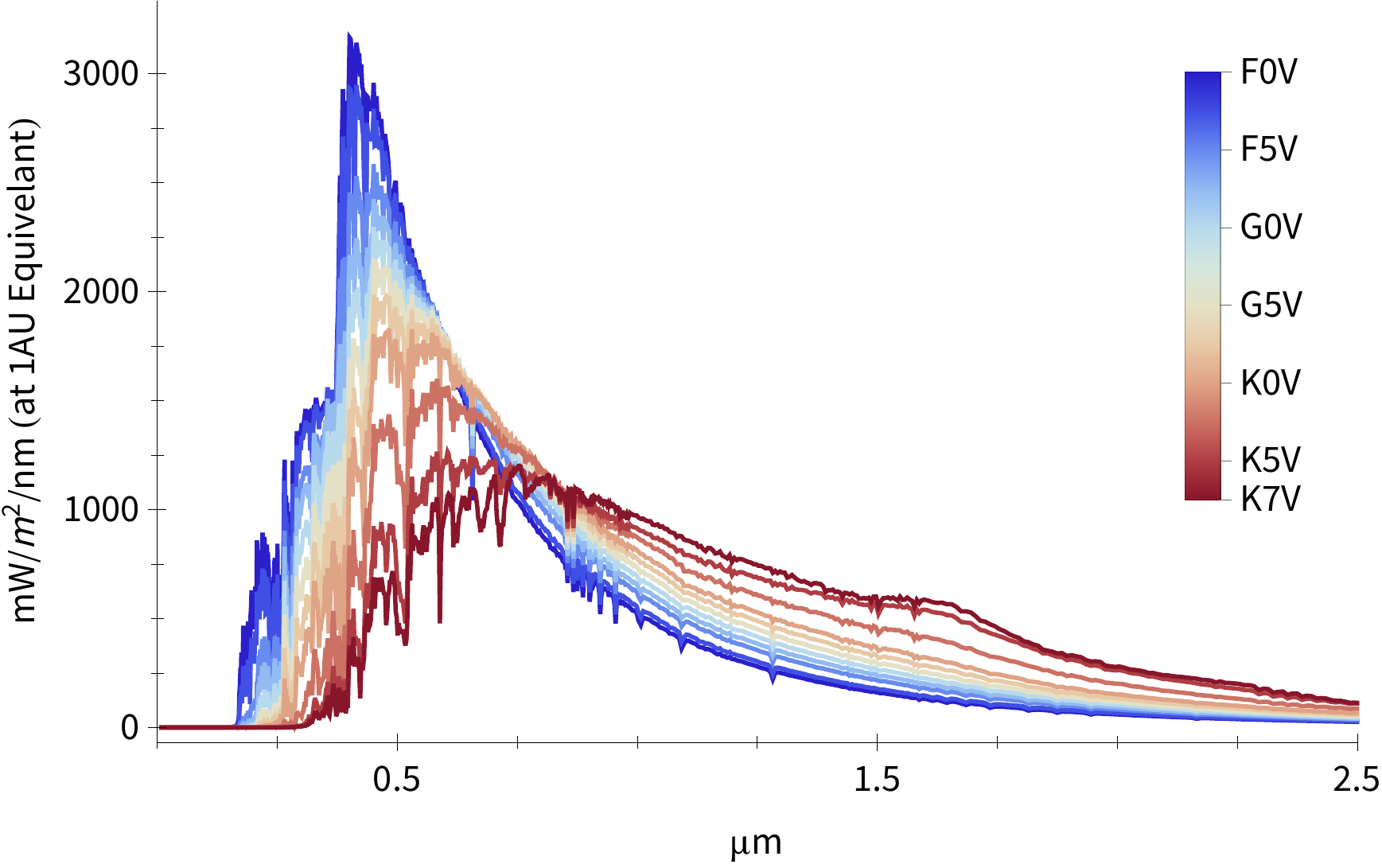}
\caption{Incident stellar flux for our model planets for host stars with $T_{eff}$ 7,400K to 3,900K, corresponding to main sequence stars F0V to K7V from ATLAS models. 
\label{fig:stellarspectra}}
\end{figure}

The effects of wavelength-dependent surface and cloud albedo on habitability are most apparent across star type. We used ATLAS models ~\citep{CastelliKurucz2004} for the F, G, and K star input spectra (Fig.  \ref{fig:stellarspectra}). We scale the spectra from these sources to provide stellar irradiation at a model planet's position, which provides similar surface temperatures for the single wavelength-independent albedo setup (as explained in section 2.1). 

\subsection{Initial Conditions}
In our models, we keep the outgassing rates for H$_2$, CH$_4$, CO, N$_2$O, and CH$_3$Cl constant, and set the mixing ratios of O$_2$ to 0.21 and CO$_2$ to $3.55\times10^{-6}$, with a varying N$_2$ concentration that is used as a fill gas to reach the set surface pressure of the model  ~\cite[see also][]{Segura2005,Segura2003,Rugheimer2013,Rugheimer2015Surface,Rugheimer2018}. Note that by keeping the outgassing rates constant, lower surface pressure atmosphere models initially have slightly higher mixing ratios of chemicals with constant outgassing ratios than higher surface pressure models. The dominant parameters we vary between simulations are the host star type and planetary surface albedo. Other parameters were altered slightly to aid in a more rapid convergence of the model, such as atmosphere height, and height of convection.

\subsection{Albedos}
The surface albedo in the 1D iterative climate-photochemistry code we updated for this study ~\citep{KastingAckerman1986,Kasting1979,Zahnle2006} was a single value from 0.237$\mu$m to 4.55$\mu$m. To simulate Earth conditions for solar irradiation at Earth's orbital position with a single albedo, a value of 0.31 is used for all wavelengths ~\cite[e.g.][]{Arney2016}. To examine how different surfaces influence a planet's climate, we updated the code to read in a wavelength-dependent albedo value.

\begin{figure}
\includegraphics[width=\columnwidth]{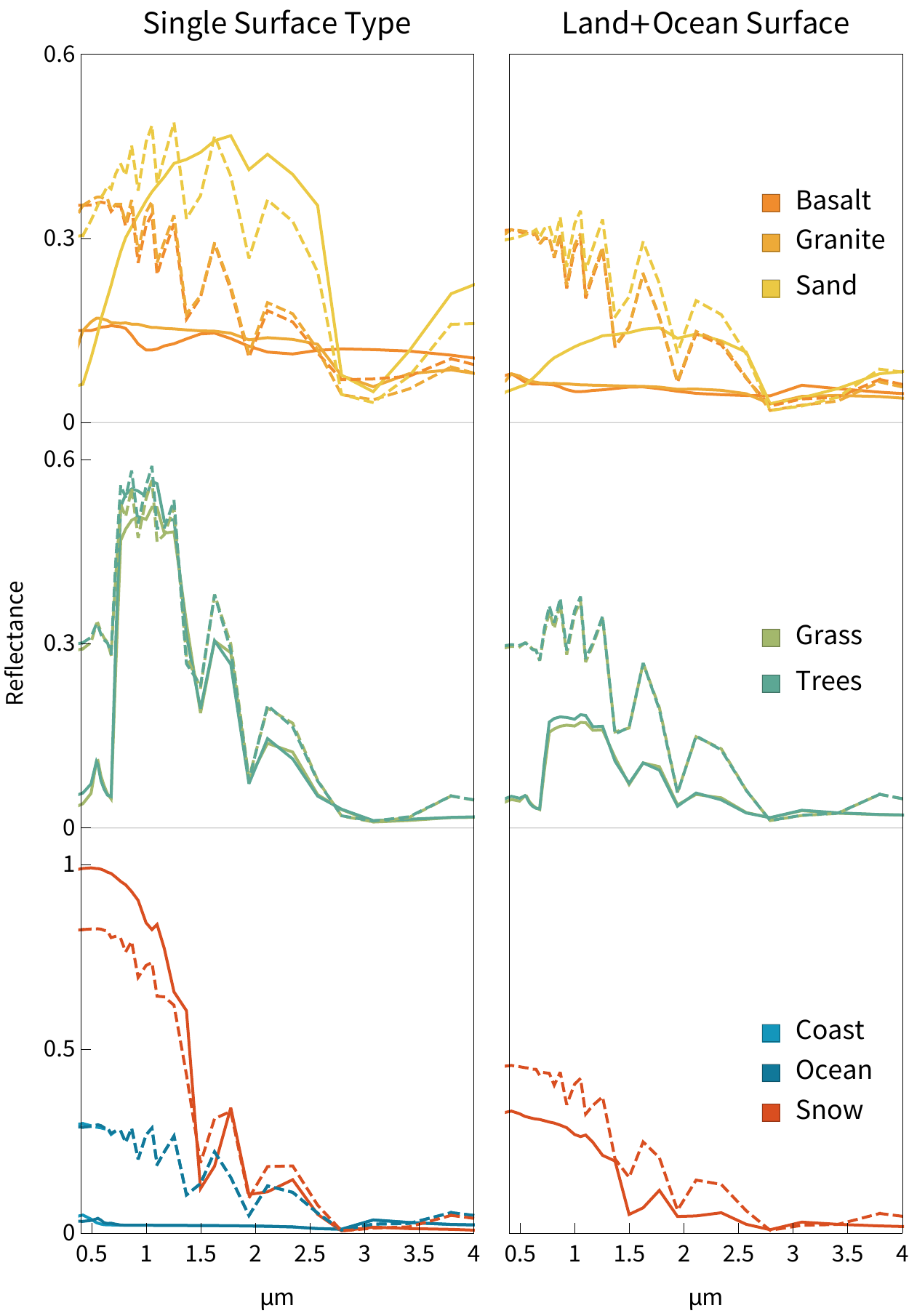}
\caption{Surface albedos used in combination to create the modern Earth surface albedo. We also use those surfaces separately to explore the climate effects of (left) a single or (right) mixed ocean-land surface albedo. Albedos are sourced from the ASTER and USGS spectral catalogs. Mixed surfaces in this figure all contain 70\% ocean and 30\% of the specified surface.\label{fig:surfacealbedos}}
\end{figure}

\begin{figure}
\includegraphics[width=\columnwidth]{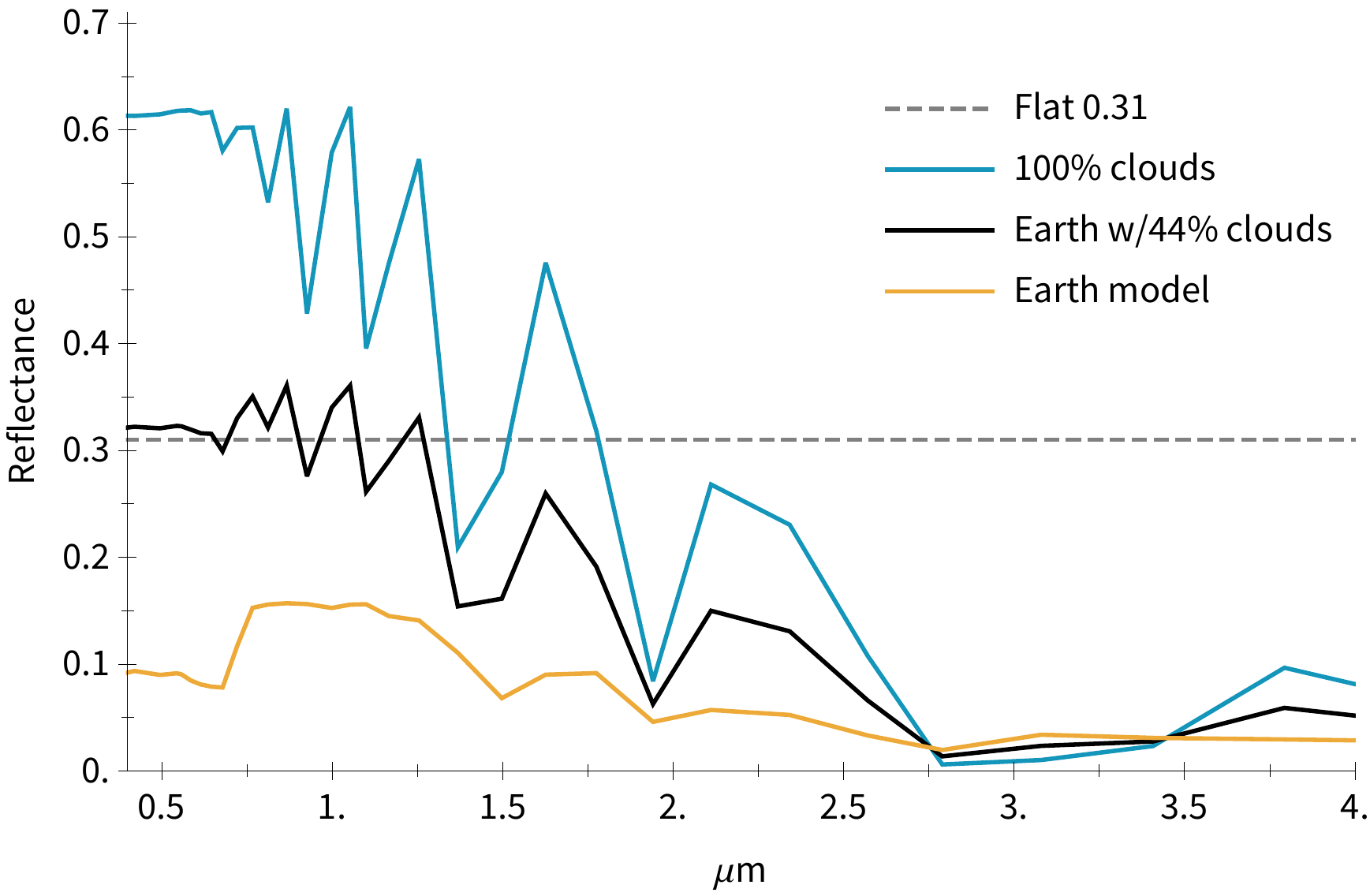}
\caption{Difference between individual Earth single-surface albedo, average Earth albedo and flat, 0.31, albedo models used in previous studies. \label{fig:earthalbedo}}
\end{figure}

Following ~\cite{Kaltenegger2007} we gathered surface albedos from the ASTER and USGS spectral libraries ~\citep{Baldridge2009aster,Kokaly2017usgs,Clark2007usgs} to create an average present-day Earth surface albedo from 8 raw albedos of snow, water, coast, sand, trees, grass, basalt, and granite (Fig. \ref{fig:surfacealbedos}). If the data was not complete in the UV, we extended the albedo constantly using the nearest value (these extensions affected regions smaller than 0.1 microns). For Earth clouds, we use the Modis 20$\mu$m cloud albedo model  ~\citep{King1997,RossowSchiffer1999}(Fig. \ref{fig:earthalbedo}), which provides an average for many clouds of different droplet size. We then used this Earth surface albedo to determine what fractional addition of clouds results in the same surface temperature as the model using a flat albedo of 0.31. 
As discussed further in the results section, we find that using 44\% cloud coverage in combination with the wavelength-dependent surface albedo of present-day Earth reproduces the same surface temperature and climate as the original albedo treatment of a single value of 0.31.

Note that the cloud fraction mimics the combined effect of warming and cooling due to clouds, which is why the fraction is lower than modern Earth's actual cloud fraction, which is between 50\% and 70\%  ~\citep{Stubenrauch2013}. Because of unknown cloud feedback for host stars with different SEDs, we then keep the cloud properties, reflectivity, and coverage constant to explore the influence of the surface albedo on the planet's climate and spectra for different host stars. 

\section{Results}
\begin{figure*}
  \begin{adjustbox}{addcode={\begin{minipage}{\width}}{\caption{Differences in model atmospheres for single-surface rocky planets orbiting in the HZ of FGK host stars with no clouds added (left) and with 44\% cloud coverage (right).
  \label{fig:profilessingle}
      }\end{minipage}},rotate=90,center}
      \includegraphics[width = \textheight]{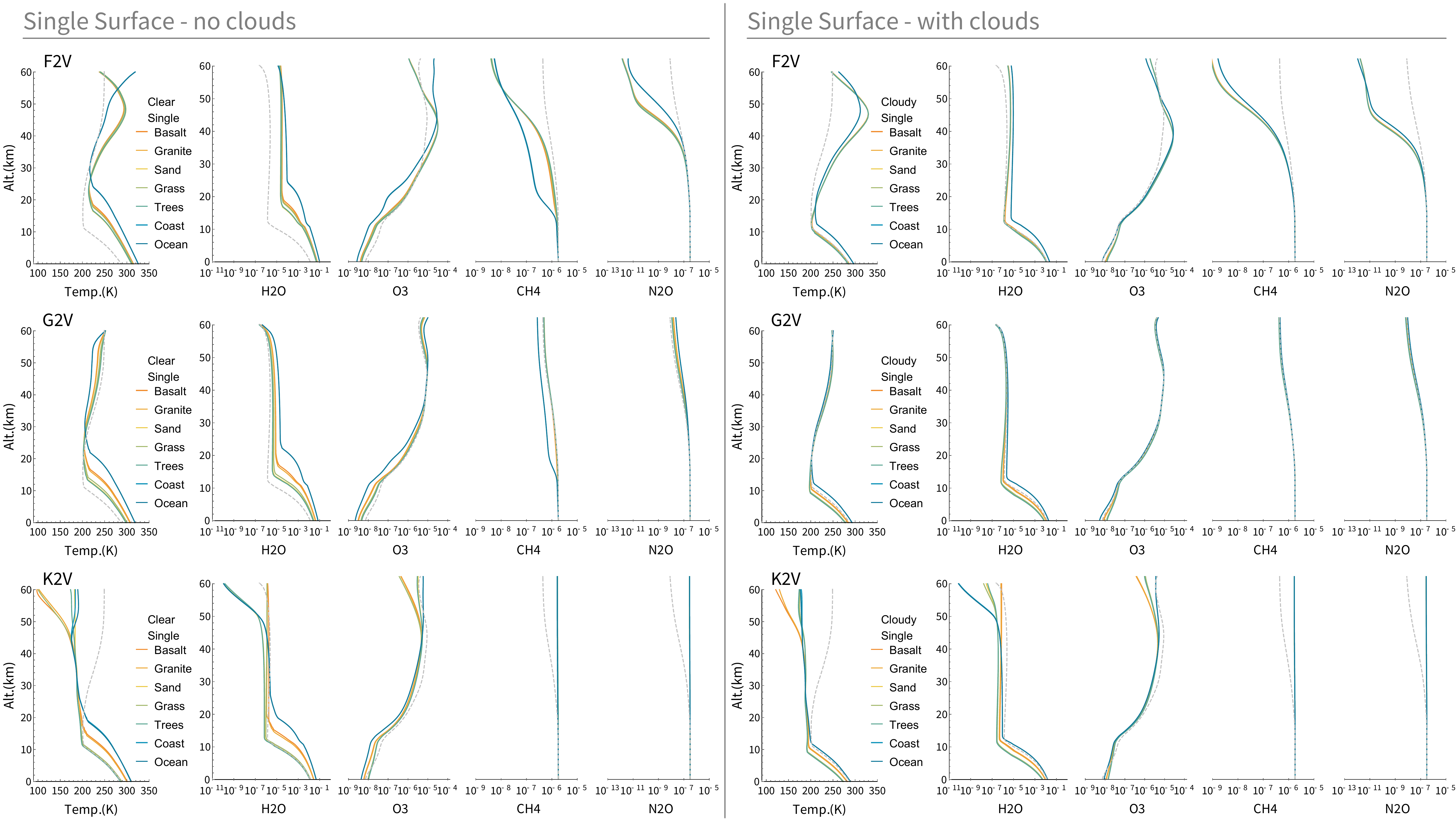}
  \end{adjustbox}
\end{figure*}

\begin{figure*}
  \begin{adjustbox}{addcode={\begin{minipage}{\width}}{\caption{Differences in model atmospheres for mixed ocean-land surface rocky planets orbiting in the HZ of FGK host stars with no clouds added (left) and with 44\% cloud coverage (right). 
  \label{fig:profilesmixed}
      }\end{minipage}},rotate=90,center}
      \includegraphics[width = \textheight]{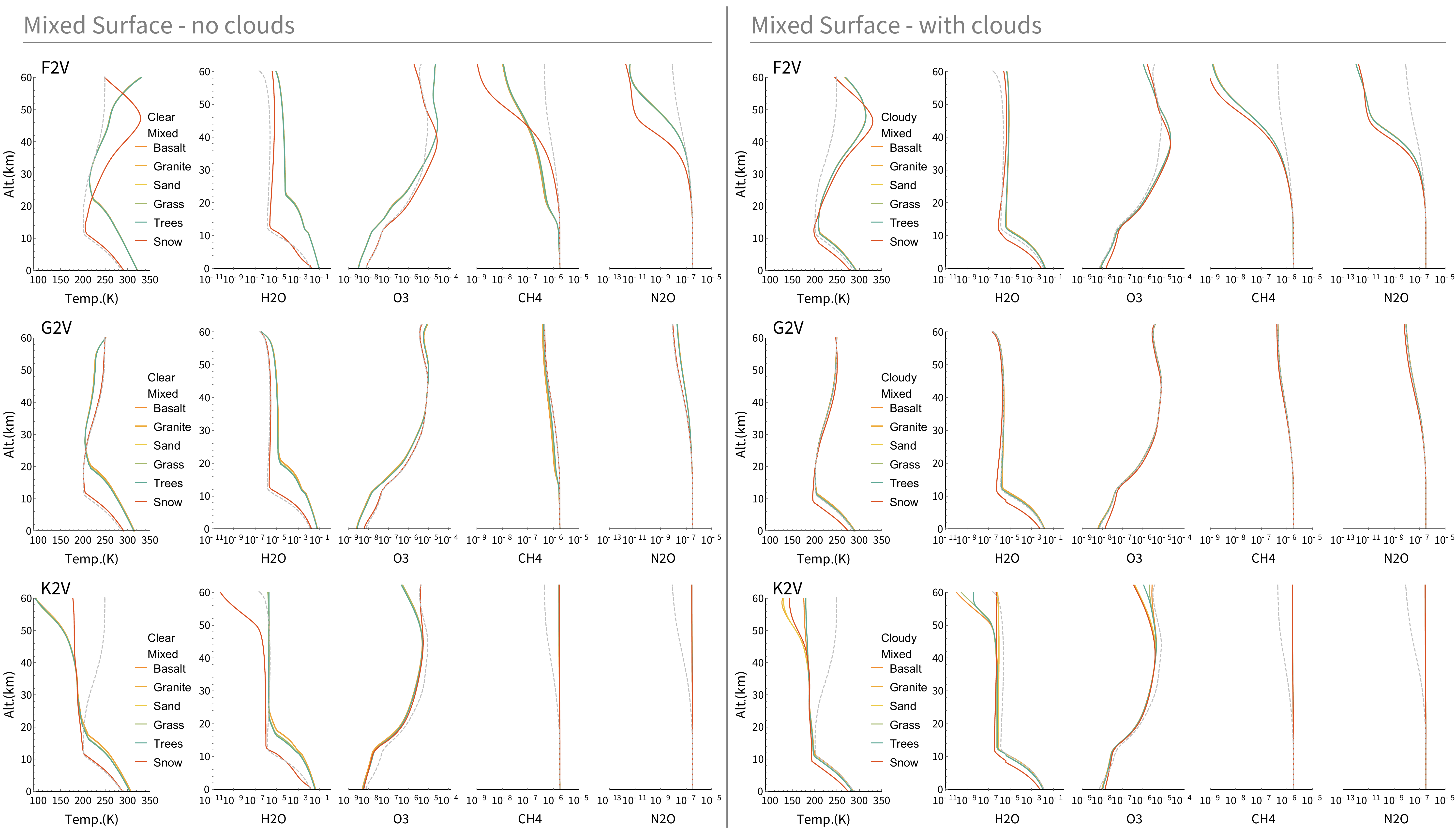}
  \end{adjustbox}
\end{figure*}

We model Earth-like planets with different surfaces for a representative grid of 12 host stars from F0 to K7 in approximately 350K effective surface temperature steps.
Figures \ref{fig:profilessingle} and \ref{fig:profilesmixed} show the temperature profiles along with the mixing ratio profiles of the major atmospheric chemicals of interest for characterization and biosignature detection H$_2$O, O$_3$, CH$_4$, and N$_2$O for planets with different surfaces with and without clouds. In this paper, we present the detailed temperature and chemical profiles of a select subset of 3 stars (K2V, G2V, F2V) and the spectra for a subset of 4 stars (K2V, G2V, F2V, F0V) in the figures for clarity. 

\subsection{Surface Temperature for Earth-like Planets}
\begin{figure}
\includegraphics[width=\columnwidth]{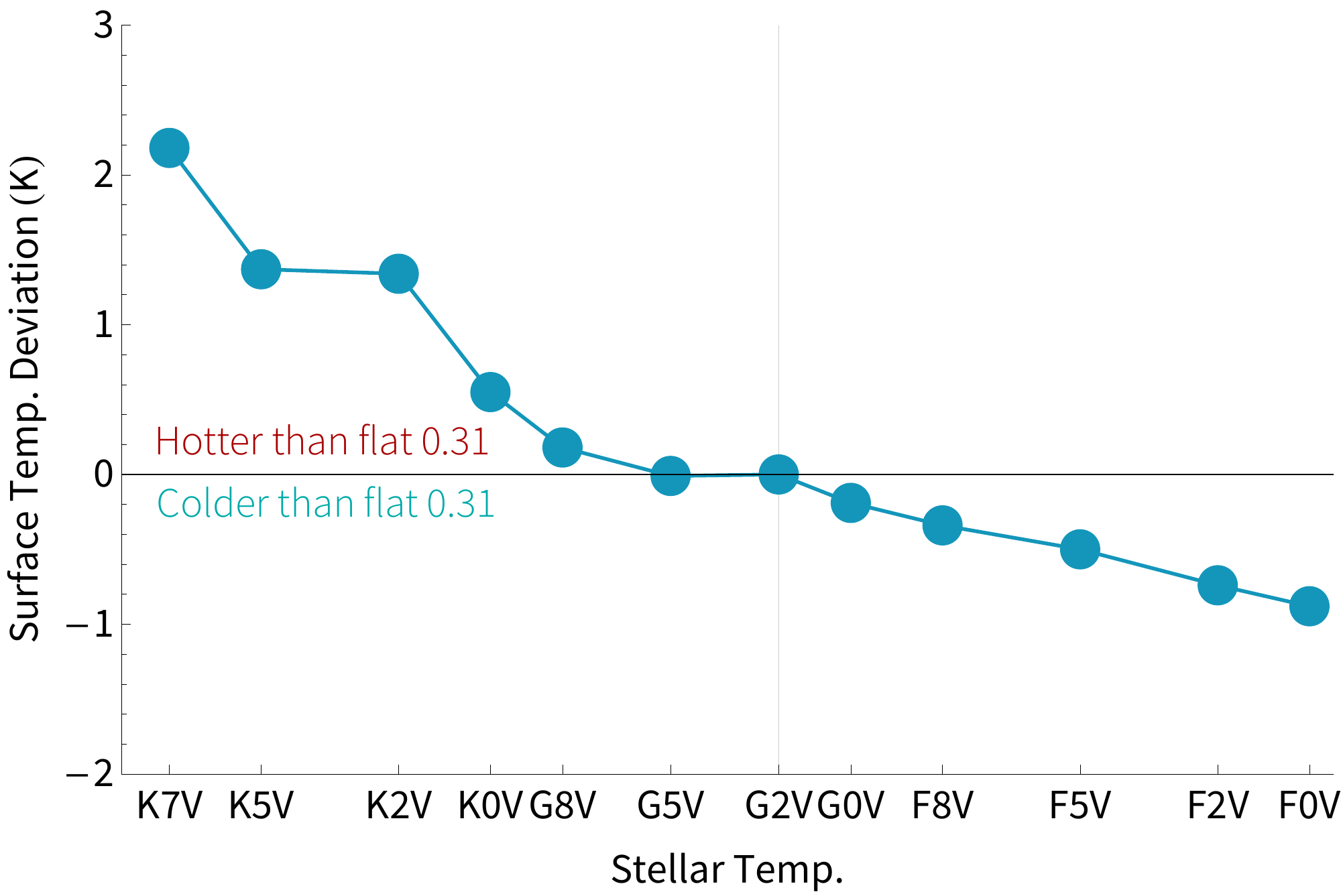}
\caption{Surface temperature deviation of model planets across host stars between using a flat 0.31 surface albedo and a wavelength-dependent Earth surface albedo with clouds. The cloud coverage in the Earth model was adjusted to create a deviation of zero for the model orbiting the Sun (G2V).   
\label{fig:surfacetemp}}
\end{figure}

Using our wavelength-dependent Earth surface albedo in combination with a 44\% cloud fraction, we compare the results for our Earth-like planet models for each host star type to the flat albedo models. Fig. \ref{fig:surfacetemp} shows that the deviation of the resulting surface temperature becomes larger as the stellar type becomes more different from the Sun. The wavelength-dependent Earth albedo is less reflective in the near IR than the flat albedo, which causes the planetary surface to become hotter around redder stars and cooler around bluer stars. For an Earth-analog surface, which the 1D flat albedo models were calibrated for, where the surface is dominated by 70\% ocean, our results show that the wavelength dependence of the surface albedo increases the average surface temperature by up to 5K and decreases it by down to 1K (Fig. \ref{fig:surfacetemp}). However, the effects can be much stronger for planets that do not have modern Earth-analog mixed surfaces.

Our models show that planetary surface temperature generally increases with decreasing effective temperature of the host star, and temperature inversions in the upper atmosphere of the planet decrease (Fig.\ref{fig:profilessingle} ~\cite[see also ][]{Segura2007, Rugheimer2015Spectra}. The absolute surface temperature generally is higher for clear atmosphere models as expected due to the high reflectivity of clouds. The surface temperature of ocean-worlds is higher due to the lower reflection of oceans compared to granite or basalt surfaces, which show higher surface temperatures than desert- and jungle-worlds.

While the trend of increasing surface temperature with decreasing host star effective temperature also holds for planets with similar surfaces, different surfaces can reduce the magnitude significantly e.g. an ocean-planet orbiting an F-star shows hotter surface temperature than a rocky planet orbiting a K-star (see Fig.\ref{fig:profilessingle}).

The chemical profiles in Figures \ref{fig:profilessingle} and \ref{fig:profilesmixed} show an increase in methane and N$_2$O concentrations as expected for host stars with lower surface temperatures, especially in the upper atmosphere ~\cite[see also][]{Segura2007, Rugheimer2015Spectra}. The ozone concentration increases for host stars with lower surface temperatures as discussed in several papers  ~\cite[e.g.][]{Segura2005,Segura2007,Rugheimer2015Spectra,Rugheimer2013}. 

\subsection{Surface Temperature of Water-, to Desert-worlds around different host stars}
\begin{figure*}
\centering
\includegraphics[height=8.8in]{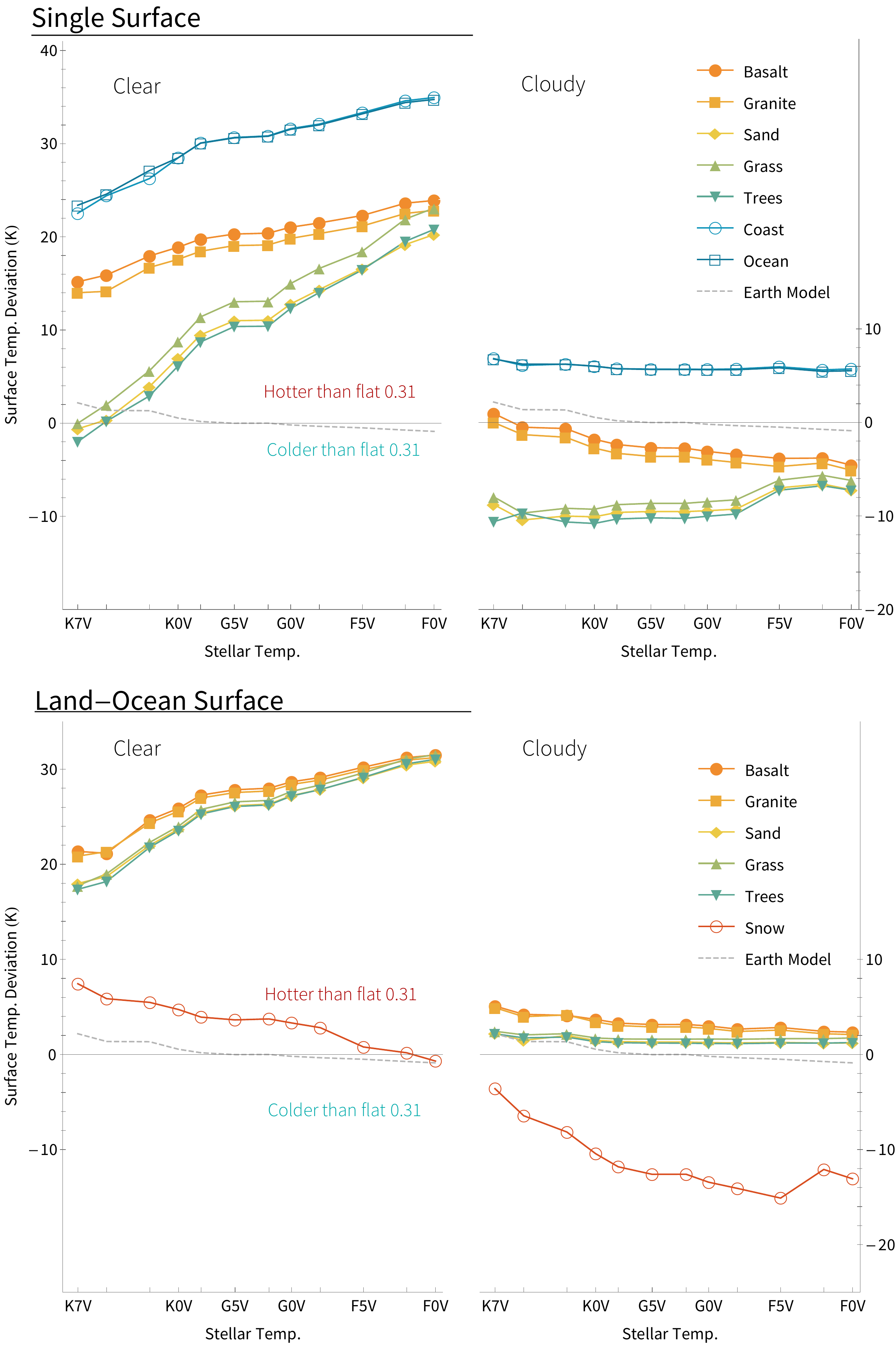} 
\caption{Planetary surface temperature deviations between models using a flat 0.31 albedo and wavelength-dependent single-surface-cloud (top) and land-ocean-cloud (bottom) albedo for different star types. Clear denotes atmosphere without and Cloudy atmospheres with 44\% Earth cloud coverage, as discussed in the text.
\label{fig:surfacetempclearcloudy}}
\end{figure*}

To explore how the spectral type of the host star influences different kinds of potentially habitable worlds, we first model a planet covered entirely with a single surface, e.g. ocean covered water worlds (using the surface albedo of oceans), desert worlds (sand), jungle worlds (trees and grass) and rocky worlds (basalt and granite) with and without clouds. Since these single-surface planets have unique, non-flat surface albedos, their climates each respond differently to host stars with different SEDs. As a second step, we created a set of wavelength-dependent ocean-land surface models comprising of the wavelength-dependent albedo of one unique surface combined with 70\% ocean coverage. Fig. \ref{fig:surfacetempclearcloudy} shows the average surface temperature differences between models assuming a flat 0.31 albedo and water-, jungle-, rocky- and desert-planets (top panel) and land-ocean surface coverage for clear and cloudy atmospheres (bottom panel). 
These results show that the surface of a planet can have a significant impact on the surface temperature of an exoplanet, and potentially alter habitability. For our single surface models, using a wavelength-dependent albedo instead of the constant albedo value changes the surface temperature for up to +35 K for an ocean planet orbiting an F0 host star and -10K for a cloudy jungle-planet orbiting a K7 host star (Fig. \ref{fig:surfacetempclearcloudy}).

\subsection{How Surfaces influence Spectra}
We model the reflection spectra of Earth-like planets with different surfaces in the Habitable Zone to explore how different surfaces influence the spectra of potentially habitable worlds seen directly imaged.

\begin{figure*}
\includegraphics[height = 4.15in]{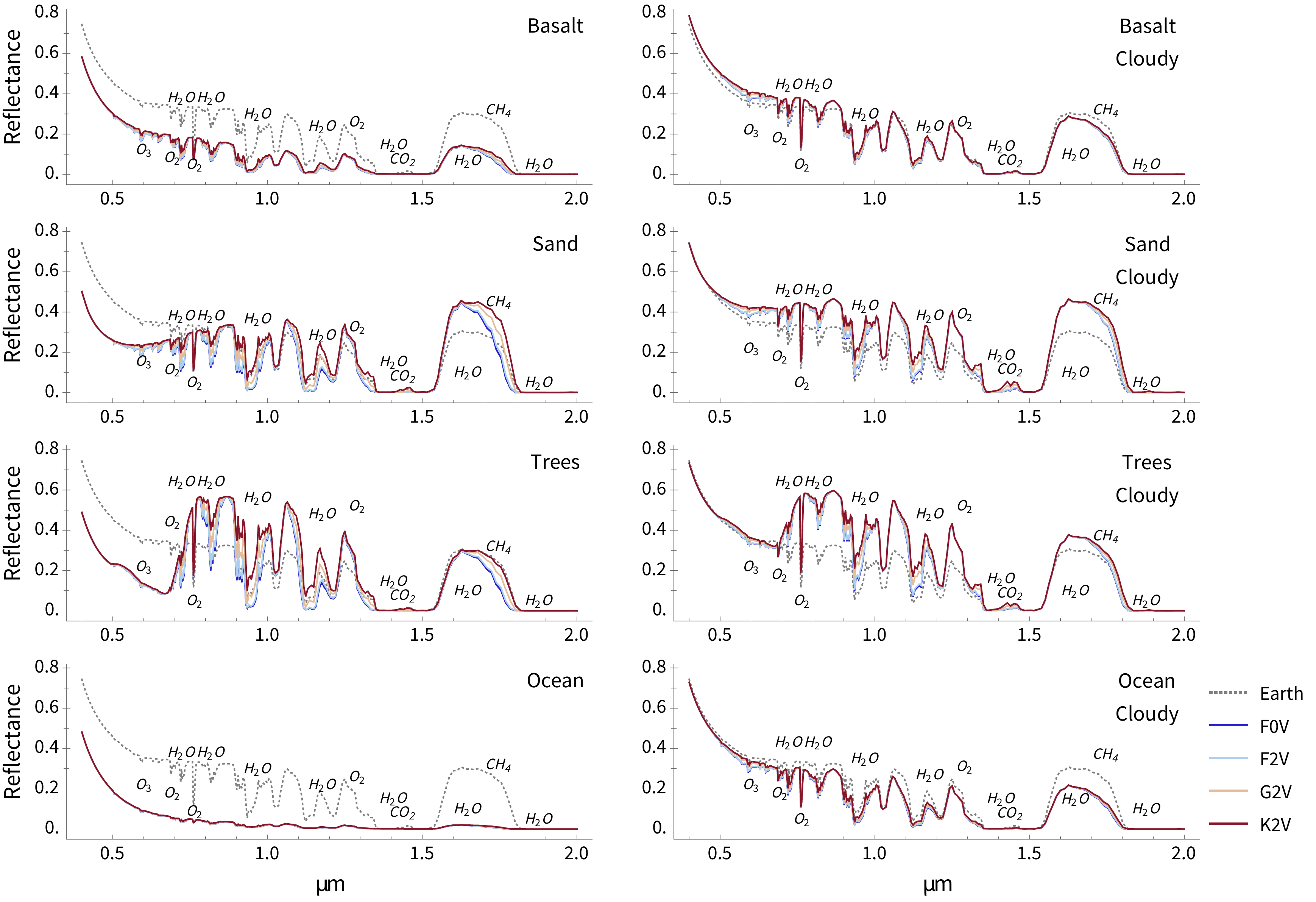}
\caption{Reflection spectra for rocky planet models with different surfaces in the HZ of F0V to K2V host stars, assuming a single-surface coverage for (left) clear and (right) cloudy atmospheres.}
\label{fig:spectrasingle}
\end{figure*}

\begin{figure*}
\includegraphics[height = 4.15in]{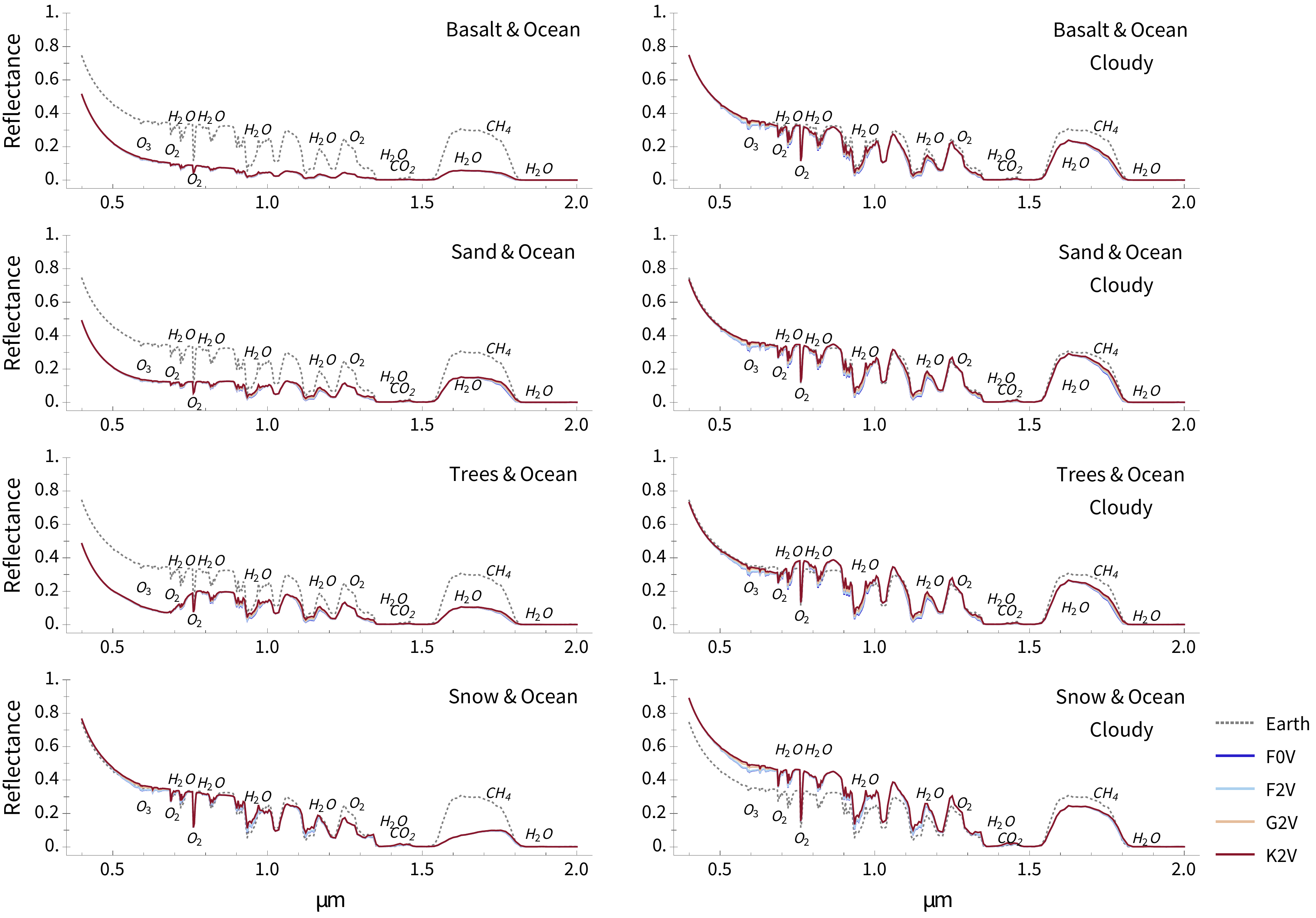}
\caption{Reflection spectra for planet models with different surfaces for rocky planets in the HZ of F0V to K2V host stars, assuming 30\% surface and 70\% ocean coverage for (left) clear and (right) cloudy atmospheres.}
\label{fig:spectramixed}
\end{figure*}

Fig. \ref{fig:spectrasingle} and Fig.\ref{fig:spectramixed} show that some atmospheric species exhibit noticeable spectral features in reflected light (0.4 to 2$\mu$m) as a result directly or indirectly from biological activity: the main ones are O$_2$, O$_3$, CH$_4$, N$_2$O and CH$_3$Cl \\
~\cite[see e.g.][]{DesMarais2002, Kaltenegger2017}. 
In the visible wavelength range, the strongest O$_2$ feature is the saturated Frauenhofer A-band at 0.76$\mu$m, with a weaker feature at 0.69$\mu$m. O$_3$ has a broad feature, the Chappuis band, which appears as a shallow triangular dip in the middle of the visible spectrum from about 0.45$\mu$m to 0.74$\mu$m. Methane at present terrestrial abundance (1.65ppm) has no significant visible absorption features, but at high abundance, it shows bands at 0.88$\mu$m, and 1.04$\mu$m, detectable e.g. in early Earth models \citep{Kaltenegger2007, Rugheimer2018}. In addition to biosignatures, H$_2$O shows bands at 0.73$\mu$m, 0.82$\mu$m, 0.95$\mu$m, and 1.14$\mu$m. CO$_2$ has negligible visible features at present abundance, but in a high CO$_2$-atmosphere of 10\% CO$_2$, like in early Earth evolution stages, the weak 1.06$\mu$m band could become detectable. 

As discussed in detail in ~\citep{Rugheimer2013,Rugheimer2015Spectra}, features of Oxygen, water, Methane, and Carbon Dioxide are present in the visible/near IR spectrum. Aside from the dominant water features, the O$_2$ feature near 0.76$\mu$ m is clearly distinguished in the relative reflection around all model stars. Most features appear deeper for the models using the wavelength-dependent Earth albedo compared to the flat albedo. Note that the spectra have not been multiplied by the incident stellar flux, which will reduce the reflected flux in the shorter wavelength range significantly ~\cite[see e.g.][]{Rugheimer2015Spectra}.

\begin{figure}
\includegraphics[width=\columnwidth]{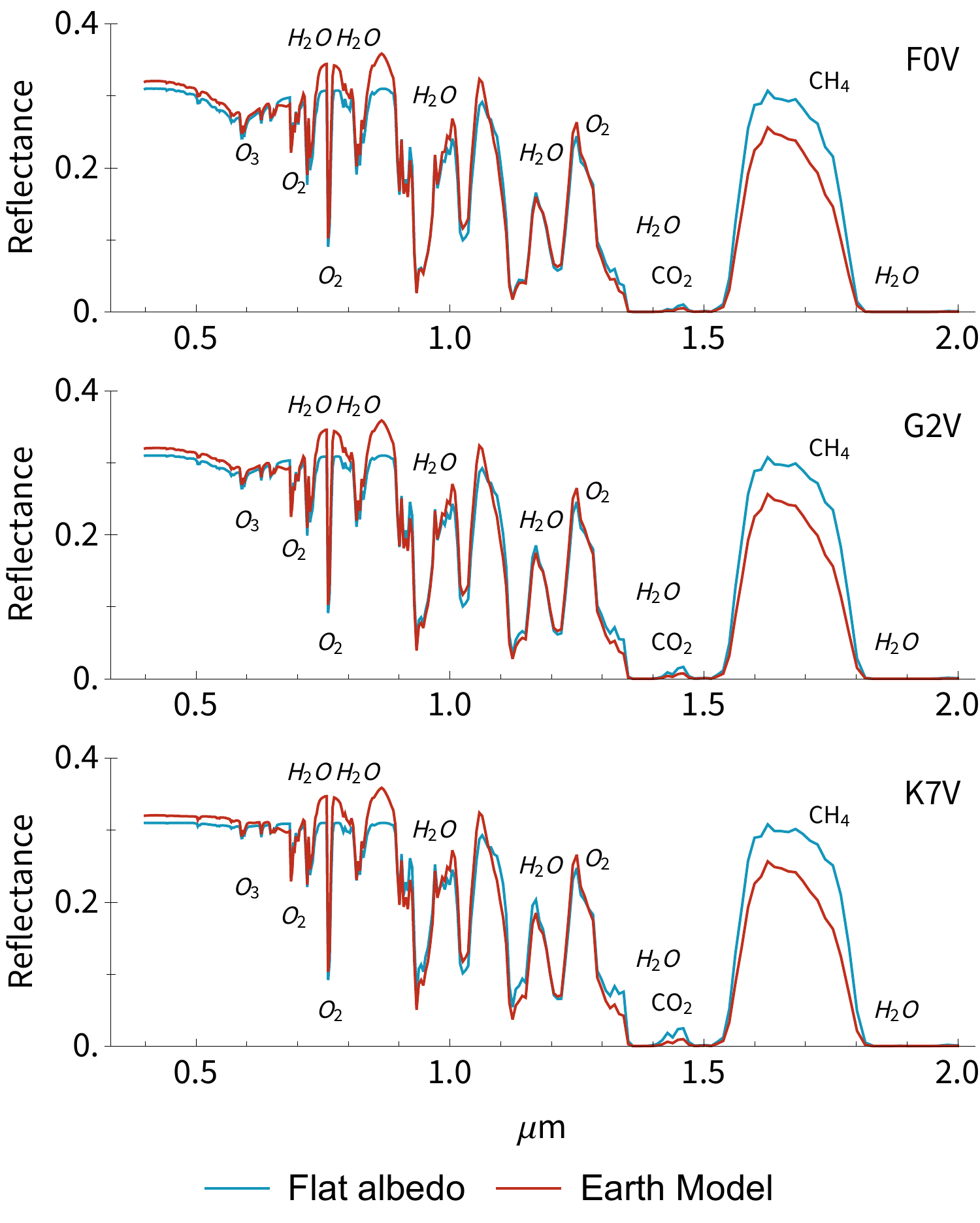}
\caption{Reflectance spectra differences for Earth models assuming a flat 0.31 albedo and a wavelength-dependent Earth albedo.
\label{fig:spectraearth}}
\end{figure}

Along with the differences in surface temperature and photochemical differences described previously, the resulting reflectance spectra show changes between a wavelength-dependent Earth-analog surface model versus a flat albedo model (Fig. \ref{fig:spectraearth}). This shows that assuming a flat albedo can both under- or over-estimate the strength of specific chemical atmospheric signatures for Earth-like planets, depending on their host stars. 

Since the wavelength-dependent Earth albedo reflects less than the flat albedo in the near IR, spectral features in the NIR will be overestimated if a flat albedo is used in an Earth spectrum model (Fig. \ref{fig:earthalbedo}).

Fig. \ref{fig:spectrasingle} and Fig. \ref{fig:spectramixed} show relative reflectance for planet models with one surface as well as 30\% land and 70\% ocean coverage, respectively, for different host stars for clear (left) and cloudy (right) atmospheres. Note that the dashed line in both figures is the modern Earth reflection spectra, including clouds. Surfaces with lower albedo (e.g. water) reduce the reflected flux of the planet. In contrast, surfaces like sand and trees increase the overall reflectivity of a planet compared to modern Earth (dashed line). 

The overall effect of the decreased stellar incident flux at shorter wavelengths for cooler stars makes atmospheric features like the 0.76$\mu$m O$_2$ challenging to detect for cooler stars ~\cite[see also e.g.][]{Rugheimer2015Spectra}. However, the shape of the surface albedo (e.g. sand reflection increases with decreasing wavelength compared to modern Earth) can increase and decrease the detectability of specific atmospheric features.

For atmospheric models, including clouds, the overall reflectivity of the planet increases due to the added high cloud albedo. However, the overall results from the single surface planet models discussed above hold, even though the coverage of the surface reduces due to the added 44\% cloud coverage in the planet models.
\\
Earth`s surface consists of about 70\% ocean and 30\% land. Therefore we also model this specific case to provide a second comparison set. Assuming one single surface dominates the remaining landmass (basalt, sand, trees, or snow), we explore the effect on the reflection spectra of such planets. Note that the 100\% ocean covered surface has been shown in Fig. \ref{fig:spectrasingle}. Adding 70\% ocean reduces the overall reflectivity of the planet because oceans only reflect a small amount of incident light (Fig. \ref{fig:surfacealbedos}). While the overall reflected flux reduces for all models (Fig. \ref{fig:spectramixed}), the overall signature changes discussed above hold for the land-ocean models as well under both clear and cloudy conditions. 

The spectra shown in Fig.\ref{fig:spectrasingle} and Fig.\ref{fig:spectramixed} show that increasing cloud coverage and decreasing surface coverage of individual surfaces decrease the detectable differences in reflected flux for planets with different surfaces (as discussed in ~\cite{Kaltenegger2007}.  

Several teams has discussed observations of Earth-size planets in the habitable zone with upcoming extremely large telescopes ~\cite[see e.g.][]{KalteneggerTraub2009,Kaltenegger2010,Stevenson2016,Barstow2016,Hedelt2013,Munoz2012,Snellen2013,Rodler2014,Betremieux2014,Misra2014}.
Different surfaces of planets can influence the overall abundance of chemicals in the atmosphere, via their influence on the surface temperature of a planet and the resulting photochemical changes. Note that these changes affect most spectral lines (especially water, oxygen, and methane).

\section{Discussion}

\subsection{Single surface models explore the most extreme effect}
Our exploration of the influence of the surface on an exoplanet`s climate and spectra shows the most extreme effect for planets fully covered with a single surface (Fig. \ref{fig:spectrasingle}). We chose this case to explore whether wavelength-dependent surfaces affected the climate and surface temperature of exoplanet models. In addition to the most extreme case, a single surface coverage, we also show a second comparison set in this paper, analog to Earth, we assume a 70\% ocean coverage of the planet`s surface. While the fraction of different surfaces on exoplanets in unknown, these two cases show i) the maximum effect of a surface on a planet`s climate as well as ii) the reduced effect if only 30\% of a planet`s surface is covered with a specific surface. While many other options are possible, these two cases clearly show how the surface fraction, as well as clouds, influence the climate of Earth-like planets. Any other surface fraction can be interpolated from the cases shown. While we do not know what surfaces exist on Earth-like exoplanets, We chose the 8 dominant surfaces of our own planet here as the test cases in our models (Fig. \ref{fig:earthalbedo}). In an exoplanet context, the interaction of wavelength-dependent surface albedo and stellar spectra has been examined mainly regarding the effect of ice-albedo feedback or special case surfaces ~\citep{Abe2011,Shields2013,Shields2014,Shields2018}. Some of our cases resembled those previously studied with snow/ice surfaces, which we confirmed show increased heating for planets around K-stars compared to G- and F-stars.

\subsection{Cloud feedback is unknown for different host stars}
Cloud feedback is unknown for different host stars. We, therefore, use clouds with similar properties for all our models based on the modern Earth model, as explained in section 2 as a first-order approximation. Cloud properties and coverage depend on many factors, including planetary rotation rate, atmospheric pressure, temperature, aerosol abundance, particle size distribution, and humidity ~\cite[see e.g.][]{Zsom2012}. Cloud coverage on Earth changes seasonally and is thought to have also changed over geologic time  ~\cite[see e.g.][]{OMalleyJames2018,Brierley2009}.

\subsection{3D and 1D model exploration of rocky planets}
3D models are being expanded to explore rocky exoplanets, with a large body of papers advancing a lively discussion in the literature on how to best include the feedback effects of clouds, rotation, surface features, atmospheric dynamics, and full photochemistry in 3D models (see e.g. discussions in \cite{GOMEZLEAL2016,GOMEZLEAL2019,Forget1997,chen2019,Lorenz1997,Williams2002,Joshi2003,Lopez2005,Selsis2007,Edson2011,Zsom2012,Goldblatt2013,Leconte2013,Leconte2013nat,Leconte2015,Vladilo2013,Wordsworth2013,Yang2013,Boschi2013,Ferreira2014,Wolf2015,Linsenmeier2015,Kopparapu2016,Popp2016,Kitzmann2017}.) 3D models and 1D models tend to be in agreement on globally averaged surface temperature for a range of stellar types though some 3D effects can exacerbate this difference due to cloud feedback \cite[][]{Arney2016}, tidal locking \cite[][]{Kopparapu2016}, or at climate extremes such as habitable zone edges \cite[][]{Shields2013,Yang2016,GOMEZLEAL2019}.

Our 1D model includes detailed photochemistry based on Earth's atmosphere and is, therefore, an excellent tool to explore the effect of the host's SED and the wavelength-dependent surfaces on the changes in the atmosphere of Earth-like planets orbiting different host stars. We concentrate on F, G, and K stars in this paper, where rocky planets in the HZ should not be synchronously locked to their host stars, and thus effective heat transfer in an Earth-like atmosphere can be assumed ~\cite[see e.g.][]{Joshi2003}. We use a 1D model for this study to explore a large range of surfaces and star types.

\section{Conclusions}

Our paper demonstrates the importance of including the wavelength-dependent feedback between a planet's surface and a planet's host star for Earth-like planets in the HZ of stars with an effective temperature between 3,900 and 7,400 K, corresponding to K7V to F0V main sequence stars. The heating or cooling effect of a specific surface is due to the interplay between the host star's SED compared to the shape of the wavelength-dependent surface albedo and can substantially change the surface temperature of an Earth-like planet. Our paper demonstrates the importance of including the wavelength-dependent feedback between a planet's surface and a planet's host star for Earth-like planets in the habitable zone. Reflected light from the surface plays a significant role not only on the overall climate but also on the detectable spectra of Earth-like planets.

\section*{Acknowledgements}

This work was supported by the Brinson Foundation and the Carl Sagan Institute. We would like to thank Zifan Lin, Thea Kozakis, and Sarah Rugheimer for comments and insights.




\bibliographystyle{mnras}

\begin{thebibliography}{}
\makeatletter
\relax
\def\mn@urlcharsother{\let\do\@makeother \do\$\do\&\do\#\do\^\do\_\do\%\do\~}
\def\mn@doi{\begingroup\mn@urlcharsother \@ifnextchar [ {\mn@doi@}
  {\mn@doi@[]}}
\def\mn@doi@[#1]#2{\def\@tempa{#1}\ifx\@tempa\@empty \href
  {http://dx.doi.org/#2} {doi:#2}\else \href {http://dx.doi.org/#2} {#1}\fi
  \endgroup}
\def\mn@eprint#1#2{\mn@eprint@#1:#2::\@nil}
\def\mn@eprint@arXiv#1{\href {http://arxiv.org/abs/#1} {{\tt arXiv:#1}}}
\def\mn@eprint@dblp#1{\href {http://dblp.uni-trier.de/rec/bibtex/#1.xml}
  {dblp:#1}}
\def\mn@eprint@#1:#2:#3:#4\@nil{\def\@tempa {#1}\def\@tempb {#2}\def\@tempc
  {#3}\ifx \@tempc \@empty \let \@tempc \@tempb \let \@tempb \@tempa \fi \ifx
  \@tempb \@empty \def\@tempb {arXiv}\fi \@ifundefined
  {mn@eprint@\@tempb}{\@tempb:\@tempc}{\expandafter \expandafter \csname
  mn@eprint@\@tempb\endcsname \expandafter{\@tempc}}}

\bibitem[\protect\citeauthoryear{{Abbot}, {Bloch-Johnson}, {Checlair},
  {Farahat}, {Graham}, {Plotkin}, {Popovic}  \& {Spaulding-Astudillo}}{{Abbot}
  et~al.}{2018}]{Abbot2018}
{Abbot} D.~S.,  {Bloch-Johnson} J.,  {Checlair} J.,  {Farahat} N.~X.,  {Graham}
  R.~J.,  {Plotkin} D.,  {Popovic} P.,   {Spaulding-Astudillo} F.,  2018,
  \mn@doi [\apj] {10.3847/1538-4357/aaa70f}, \href
  {https://ui.adsabs.harvard.edu/abs/2018ApJ...854....3A} {854, 3}

\bibitem[\protect\citeauthoryear{{Abe}, {Abe-Ouchi}, {Sleep}  \&
  {Zahnle}}{{Abe} et~al.}{2011}]{Abe2011}
{Abe} Y.,  {Abe-Ouchi} A.,  {Sleep} N.~H.,   {Zahnle} K.~J.,  2011, \mn@doi
  [Astrobiology] {10.1089/ast.2010.0545}, \href
  {https://ui.adsabs.harvard.edu/abs/2011AsBio..11..443A} {11, 443}

\bibitem[\protect\citeauthoryear{{Airapetian}, {Glocer}, {Khazanov}, {Loyd},
  {France}, {Sojka}, {Danchi}  \& {Liemohn}}{{Airapetian}
  et~al.}{2017}]{Airapetian2017}
{Airapetian} V.~S.,  {Glocer} A.,  {Khazanov} G.~V.,  {Loyd} R.~O.~P.,
  {France} K.,  {Sojka} J.,  {Danchi} W.~C.,   {Liemohn} M.~W.,  2017, \mn@doi
  [\apjl] {10.3847/2041-8213/836/1/L3}, \href
  {https://ui.adsabs.harvard.edu/abs/2017ApJ...836L...3A} {836, L3}

\bibitem[\protect\citeauthoryear{{Arney} et~al.,}{{Arney}
  et~al.}{2016}]{Arney2016}
{Arney} G.,  et~al., 2016, \mn@doi [Astrobiology] {10.1089/ast.2015.1422},
  \href {https://ui.adsabs.harvard.edu/abs/2016AsBio..16..873A} {16, 873}

\bibitem[\protect\citeauthoryear{Baldridge, Hook, Grove  \& Rivera}{Baldridge
  et~al.}{2009}]{Baldridge2009aster}
Baldridge A.~M.,  Hook S.,  Grove C.,   Rivera G.,  2009, Remote Sensing of
  Environment, 113, 711

\bibitem[\protect\citeauthoryear{{Barstow} \& {Irwin}}{{Barstow} \&
  {Irwin}}{2016}]{Barstow2016}
{Barstow} J.~K.,  {Irwin} P.~G.~J.,  2016, \mn@doi [Monthly Notices of the
  Royal Astronomical Society] {10.1093/mnrasl/slw109}, \href
  {https://ui.adsabs.harvard.edu/abs/2016MNRAS.461L..92B} {461, L92}

\bibitem[\protect\citeauthoryear{{Battersby} et~al.,}{{Battersby}
  et~al.}{2018}]{Battersby2018}
{Battersby} C.,  et~al., 2018, \mn@doi [Nature Astronomy]
  {10.1038/s41550-018-0540-y}, \href
  {https://ui.adsabs.harvard.edu/abs/2018NatAs...2..596B} {2, 596}

\bibitem[\protect\citeauthoryear{{B{\'e}tr{\'e}mieux} \&
  {Kaltenegger}}{{B{\'e}tr{\'e}mieux} \& {Kaltenegger}}{2014}]{Betremieux2014}
{B{\'e}tr{\'e}mieux} Y.,  {Kaltenegger} L.,  2014, \mn@doi [\apj]
  {10.1088/0004-637X/791/1/7}, \href
  {https://ui.adsabs.harvard.edu/abs/2014ApJ...791....7B} {791, 7}

\bibitem[\protect\citeauthoryear{{Boschi}, {Lucarini}  \& {Pascale}}{{Boschi}
  et~al.}{2013}]{Boschi2013}
{Boschi} R.,  {Lucarini} V.,   {Pascale} S.,  2013, \mn@doi [\icarus]
  {10.1016/j.icarus.2013.03.017}, \href
  {https://ui.adsabs.harvard.edu/abs/2013Icar..226.1724B} {226, 1724}

\bibitem[\protect\citeauthoryear{Brierley, Fedorov, Liu, Herbert, Lawrence  \&
  LaRiviere}{Brierley et~al.}{2009}]{Brierley2009}
Brierley C.~M.,  Fedorov A.~V.,  Liu Z.,  Herbert T.~D.,  Lawrence K.~T.,
  LaRiviere J.~P.,  2009, \mn@doi [Science] {10.1126/science.1167625}, 323,
  1714

\bibitem[\protect\citeauthoryear{{Castelli} \& {Kurucz}}{{Castelli} \&
  {Kurucz}}{2004}]{CastelliKurucz2004}
{Castelli} F.,  {Kurucz} R.~L.,  2004, ArXiv Astrophysics e-prints, \href
  {http://adsabs.harvard.edu/abs/2004astro.ph..5087C} {}

\bibitem[\protect\citeauthoryear{{Chen}, {Wolf}, {Zhan}  \& {Horton}}{{Chen}
  et~al.}{2019}]{chen2019}
{Chen} H.,  {Wolf} E.~T.,  {Zhan} Z.,   {Horton} D.~E.,  2019, \mn@doi [\apj]
  {10.3847/1538-4357/ab4f7e}, \href
  {https://ui.adsabs.harvard.edu/abs/2019ApJ...886...16C} {886, 16}

\bibitem[\protect\citeauthoryear{Clark, Swayze, Wise, Livo, Hoefen, Kokaly  \&
  Sutley}{Clark et~al.}{2007}]{Clark2007usgs}
Clark R.~N.,  Swayze G.~A.,  Wise R.~A.,  Livo K.~E.,  Hoefen T.~M.,  Kokaly
  R.~F.,   Sutley S.~J.,  2007, Technical report, USGS digital spectral library
  splib06a.
US Geological Survey

\bibitem[\protect\citeauthoryear{{Cockell}, {Kaltenegger}  \&
  {Raven}}{{Cockell} et~al.}{2009}]{Cockell2009}
{Cockell} C.~S.,  {Kaltenegger} L.,   {Raven} J.~A.,  2009, \mn@doi
  [Astrobiology] {10.1089/ast.2008.0273}, \href
  {https://ui.adsabs.harvard.edu/abs/2009AsBio...9..623C} {9, 623}

\bibitem[\protect\citeauthoryear{{Des Marais} et~al.,}{{Des Marais}
  et~al.}{2002}]{DesMarais2002}
{Des Marais} D.~J.,  et~al., 2002, \mn@doi [Astrobiology]
  {10.1089/15311070260192246}, \href
  {https://ui.adsabs.harvard.edu/abs/2002AsBio...2..153D} {2, 153}

\bibitem[\protect\citeauthoryear{{Edson}, {Lee}, {Bannon}, {Kasting}  \&
  {Pollard}}{{Edson} et~al.}{2011}]{Edson2011}
{Edson} A.,  {Lee} S.,  {Bannon} P.,  {Kasting} J.~F.,   {Pollard} D.,  2011,
  \mn@doi [\icarus] {10.1016/j.icarus.2010.11.023}, \href
  {https://ui.adsabs.harvard.edu/abs/2011Icar..212....1E} {212, 1}

\bibitem[\protect\citeauthoryear{{Ferreira}, {Marshall}, {O'Gorman}  \&
  {Seager}}{{Ferreira} et~al.}{2014}]{Ferreira2014}
{Ferreira} D.,  {Marshall} J.,  {O'Gorman} P.~A.,   {Seager} S.,  2014, \mn@doi
  [\icarus] {10.1016/j.icarus.2014.09.015}, \href
  {https://ui.adsabs.harvard.edu/abs/2014Icar..243..236F} {243, 236}

\bibitem[\protect\citeauthoryear{{Forget} \& {Pierrehumbert}}{{Forget} \&
  {Pierrehumbert}}{1997}]{Forget1997}
{Forget} F.,  {Pierrehumbert} R.~T.,  1997, \mn@doi [Science]
  {10.1126/science.278.5341.1273}, \href
  {https://ui.adsabs.harvard.edu/abs/1997Sci...278.1273F} {278, 1273}

\bibitem[\protect\citeauthoryear{{Garc{\'\i}a Mu{\~n}oz}, {Zapatero Osorio},
  {Barrena}, {Monta{\~n}{\'e}s-Rodr{\'\i}guez}, {Mart{\'\i}n}  \&
  {Pall{\'e}}}{{Garc{\'\i}a Mu{\~n}oz} et~al.}{2012}]{Munoz2012}
{Garc{\'\i}a Mu{\~n}oz} A.,  {Zapatero Osorio} M.~R.,  {Barrena} R.,
  {Monta{\~n}{\'e}s-Rodr{\'\i}guez} P.,  {Mart{\'\i}n} E.~L.,   {Pall{\'e}} E.,
   2012, \mn@doi [\apj] {10.1088/0004-637X/755/2/103}, \href
  {https://ui.adsabs.harvard.edu/abs/2012ApJ...755..103G} {755, 103}

\bibitem[\protect\citeauthoryear{{Goldblatt}, {Robinson}, {Zahnle}  \&
  {Crisp}}{{Goldblatt} et~al.}{2013}]{Goldblatt2013}
{Goldblatt} C.,  {Robinson} T.~D.,  {Zahnle} K.~J.,   {Crisp} D.,  2013,
  \mn@doi [Nature Geoscience] {10.1038/ngeo1892}, \href
  {https://ui.adsabs.harvard.edu/abs/2013NatGe...6..661G} {6, 661}

\bibitem[\protect\citeauthoryear{Gomez-Leal, Codron  \& Selsis}{Gomez-Leal
  et~al.}{2016}]{GOMEZLEAL2016}
Gomez-Leal I.,  Codron F.,   Selsis F.,  2016, \mn@doi [Icarus]
  {https://doi.org/10.1016/j.icarus.2015.12.050}, 269, 98

\bibitem[\protect\citeauthoryear{Gomez-Leal, Kaltenegger, Lucarini  \&
  Lunkeit}{Gomez-Leal et~al.}{2019}]{GOMEZLEAL2019}
Gomez-Leal I.,  Kaltenegger L.,  Lucarini V.,   Lunkeit F.,  2019, \mn@doi
  [Icarus] {https://doi.org/10.1016/j.icarus.2018.11.019}, 321, 608

\bibitem[\protect\citeauthoryear{{Haqq-Misra}, {Domagal-Goldman}, {Kasting}  \&
  {Kasting}}{{Haqq-Misra} et~al.}{2008}]{HaqqMisra2008}
{Haqq-Misra} J.~D.,  {Domagal-Goldman} S.~D.,  {Kasting} P.~J.,   {Kasting}
  J.~F.,  2008, \mn@doi [Astrobiology] {10.1089/ast.2007.0197}, \href
  {https://ui.adsabs.harvard.edu/abs/2008AsBio...8.1127H} {8, 1127}

\bibitem[\protect\citeauthoryear{{Hedelt} et~al.,}{{Hedelt}
  et~al.}{2013}]{Hedelt2013}
{Hedelt} P.,  et~al., 2013, \mn@doi [\aap] {10.1051/0004-6361/201117723}, \href
  {https://ui.adsabs.harvard.edu/abs/2013A&A...553A...9H} {553, A9}

\bibitem[\protect\citeauthoryear{{Johnstone}, {G{\"u}del}, {Lammer}  \&
  {Kislyakova}}{{Johnstone} et~al.}{2018}]{Johnstone2018}
{Johnstone} C.~P.,  {G{\"u}del} M.,  {Lammer} H.,   {Kislyakova} K.~G.,  2018,
  \mn@doi [\aap] {10.1051/0004-6361/201832776}, \href
  {https://ui.adsabs.harvard.edu/abs/2018A&A...617A.107J} {617, A107}

\bibitem[\protect\citeauthoryear{{Joshi}}{{Joshi}}{2003}]{Joshi2003}
{Joshi} M.,  2003, \mn@doi [Astrobiology] {10.1089/153110703769016488}, \href
  {https://ui.adsabs.harvard.edu/abs/2003AsBio...3..415J} {3, 415}

\bibitem[\protect\citeauthoryear{{Kaltenegger}}{{Kaltenegger}}{2017}]{Kaltenegger2017}
{Kaltenegger} L.,  2017, \mn@doi [Annual Review of Astronomy and Astrophysics]
  {10.1146/annurev-astro-082214-122238}, \href
  {https://ui.adsabs.harvard.edu/abs/2017ARA&A..55..433K} {55, 433}

\bibitem[\protect\citeauthoryear{{Kaltenegger} \& {Haghighipour}}{{Kaltenegger}
  \& {Haghighipour}}{2013}]{Kaltenegger2013}
{Kaltenegger} L.,  {Haghighipour} N.,  2013, \mn@doi [The Astrophysical
  Journal] {10.1088/0004-637X/777/2/165}, \href
  {https://ui.adsabs.harvard.edu/abs/2013ApJ...777..165K} {777, 165}

\bibitem[\protect\citeauthoryear{{Kaltenegger} \& {Traub}}{{Kaltenegger} \&
  {Traub}}{2009}]{KalteneggerTraub2009}
{Kaltenegger} L.,  {Traub} W.~A.,  2009, \mn@doi [The Astrophysical Journal]
  {10.1088/0004-637X/698/1/519}, \href
  {https://ui.adsabs.harvard.edu/abs/2009ApJ...698..519K} {698, 519}

\bibitem[\protect\citeauthoryear{{Kaltenegger}, {Traub}  \&
  {Jucks}}{{Kaltenegger} et~al.}{2007}]{Kaltenegger2007}
{Kaltenegger} L.,  {Traub} W.~A.,   {Jucks} K.~W.,  2007, \mn@doi [The
  Astrophysical Journal] {10.1086/510996}, \href
  {https://ui.adsabs.harvard.edu/abs/2007ApJ...658..598K} {658, 598}

\bibitem[\protect\citeauthoryear{{Kaltenegger} et~al.,}{{Kaltenegger}
  et~al.}{2010}]{Kaltenegger2010}
{Kaltenegger} L.,  et~al., 2010, \mn@doi [Astrobiology]
  {10.1089/ast.2009.0381}, \href
  {https://ui.adsabs.harvard.edu/abs/2010AsBio..10...89K} {10, 89}

\bibitem[\protect\citeauthoryear{{Kane} \& {Hinkel}}{{Kane} \&
  {Hinkel}}{2013}]{Kane2013}
{Kane} S.~R.,  {Hinkel} N.~R.,  2013, \mn@doi [\apj]
  {10.1088/0004-637X/762/1/7}, \href
  {https://ui.adsabs.harvard.edu/abs/2013ApJ...762....7K} {762, 7}

\bibitem[\protect\citeauthoryear{{Kasting} \& {Ackerman}}{{Kasting} \&
  {Ackerman}}{1986}]{KastingAckerman1986}
{Kasting} J.~F.,  {Ackerman} T.~P.,  1986, \mn@doi [Science]
  {10.1126/science.234.4782.1383}, \href
  {https://ui.adsabs.harvard.edu/abs/1986Sci...234.1383K} {234, 1383}

\bibitem[\protect\citeauthoryear{{Kasting}, {Liu}  \& {Donahue}}{{Kasting}
  et~al.}{1979}]{Kasting1979}
{Kasting} J.~F.,  {Liu} S.~C.,   {Donahue} T.~M.,  1979, \mn@doi [\jgr]
  {10.1029/JC084iC06p03097}, \href
  {https://ui.adsabs.harvard.edu/abs/1979JGR....84.3097K} {84, 3097}

\bibitem[\protect\citeauthoryear{{Kasting}, {Holland}  \& {Pinto}}{{Kasting}
  et~al.}{1985}]{Kasting1985}
{Kasting} J.~F.,  {Holland} H.~D.,   {Pinto} J.~P.,  1985, \mn@doi [\jgr]
  {10.1029/JD090iD06p10497}, \href
  {https://ui.adsabs.harvard.edu/abs/1985JGR....9010497K} {90, 10,497}

\bibitem[\protect\citeauthoryear{{Kasting}, {Whitmire}  \&
  {Reynolds}}{{Kasting} et~al.}{1993}]{Kasting1993}
{Kasting} J.~F.,  {Whitmire} D.~P.,   {Reynolds} R.~T.,  1993, \mn@doi
  [\icarus] {10.1006/icar.1993.1010}, \href
  {https://ui.adsabs.harvard.edu/abs/1993Icar..101..108K} {101, 108}

\bibitem[\protect\citeauthoryear{King, Tsay, Platnick, Wang  \& Liou}{King
  et~al.}{1997}]{King1997}
King M.,  Tsay S.-C.,  Platnick S.,  Wang M.,   Liou K.,  1997, MODIS Algorithm
  Theoretical Basis Document, No. ATBD-MOD-05

\bibitem[\protect\citeauthoryear{{Kitzmann}}{{Kitzmann}}{2017}]{Kitzmann2017}
{Kitzmann} D.,  2017, \mn@doi [\aap] {10.1051/0004-6361/201630029}, \href
  {https://ui.adsabs.harvard.edu/abs/2017A&A...600A.111K} {600, A111}

\bibitem[\protect\citeauthoryear{Kokaly et~al.,}{Kokaly
  et~al.}{2017}]{Kokaly2017usgs}
Kokaly R.~F.,  et~al., 2017, Technical report, USGS spectral library version 7.
US Geological Survey

\bibitem[\protect\citeauthoryear{{Kopparapu} et~al.,}{{Kopparapu}
  et~al.}{2013}]{Kopparapu2013}
{Kopparapu} R.~K.,  et~al., 2013, \mn@doi [\apj] {10.1088/0004-637X/765/2/131},
  \href {https://ui.adsabs.harvard.edu/abs/2013ApJ...765..131K} {765, 131}

\bibitem[\protect\citeauthoryear{{Kopparapu}, {Wolf}, {Haqq-Misra}, {Yang},
  {Kasting}, {Meadows}, {Terrien}  \& {Mahadevan}}{{Kopparapu}
  et~al.}{2016}]{Kopparapu2016}
{Kopparapu} R.~k.,  {Wolf} E.~T.,  {Haqq-Misra} J.,  {Yang} J.,  {Kasting}
  J.~F.,  {Meadows} V.,  {Terrien} R.,   {Mahadevan} S.,  2016, \mn@doi [\apj]
  {10.3847/0004-637X/819/1/84}, \href
  {https://ui.adsabs.harvard.edu/abs/2016ApJ...819...84K} {819, 84}

\bibitem[\protect\citeauthoryear{{Leconte}, {Forget}, {Charnay}, {Wordsworth}
  \& {Pottier}}{{Leconte} et~al.}{2013a}]{Leconte2013nat}
{Leconte} J.,  {Forget} F.,  {Charnay} B.,  {Wordsworth} R.,   {Pottier} A.,
  2013a, \mn@doi [\nat] {10.1038/nature12827}, \href
  {https://ui.adsabs.harvard.edu/abs/2013Natur.504..268L} {504, 268}

\bibitem[\protect\citeauthoryear{{Leconte}, {Forget}, {Charnay}, {Wordsworth},
  {Selsis}, {Millour}  \& {Spiga}}{{Leconte} et~al.}{2013b}]{Leconte2013}
{Leconte} J.,  {Forget} F.,  {Charnay} B.,  {Wordsworth} R.,  {Selsis} F.,
  {Millour} E.,   {Spiga} A.,  2013b, \mn@doi [\aap]
  {10.1051/0004-6361/201321042}, \href
  {https://ui.adsabs.harvard.edu/abs/2013A&A...554A..69L} {554, A69}

\bibitem[\protect\citeauthoryear{{Leconte}, {Wu}, {Menou}  \&
  {Murray}}{{Leconte} et~al.}{2015}]{Leconte2015}
{Leconte} J.,  {Wu} H.,  {Menou} K.,   {Murray} N.,  2015, \mn@doi [Science]
  {10.1126/science.1258686}, \href
  {https://ui.adsabs.harvard.edu/abs/2015Sci...347..632L} {347, 632}

\bibitem[\protect\citeauthoryear{{Lederberg}}{{Lederberg}}{1965}]{Lederberg1965}
{Lederberg} J.,  1965, \mn@doi [\nat] {10.1038/207009a0}, \href
  {https://ui.adsabs.harvard.edu/abs/1965Natur.207....9L} {207, 9}

\bibitem[\protect\citeauthoryear{{Linsenmeier}, {Pascale}  \&
  {Lucarini}}{{Linsenmeier} et~al.}{2015}]{Linsenmeier2015}
{Linsenmeier} M.,  {Pascale} S.,   {Lucarini} V.,  2015, \mn@doi [\planss]
  {10.1016/j.pss.2014.11.003}, \href
  {https://ui.adsabs.harvard.edu/abs/2015P&SS..105...43L} {105, 43}

\bibitem[\protect\citeauthoryear{{Lippincott}, {Eck}, {Dayhoff}  \&
  {Sagan}}{{Lippincott} et~al.}{1967}]{Lippincott1967}
{Lippincott} E.~R.,  {Eck} R.~V.,  {Dayhoff} M.~O.,   {Sagan} C.,  1967,
  \mn@doi [\apj] {10.1086/149051}, \href
  {https://ui.adsabs.harvard.edu/abs/1967ApJ...147..753L} {147, 753}

\bibitem[\protect\citeauthoryear{{Lopez}, {Schneider}  \& {Danchi}}{{Lopez}
  et~al.}{2005}]{Lopez2005}
{Lopez} B.,  {Schneider} J.,   {Danchi} W.~C.,  2005, \mn@doi [\apj]
  {10.1086/430416}, \href
  {https://ui.adsabs.harvard.edu/abs/2005ApJ...627..974L} {627, 974}

\bibitem[\protect\citeauthoryear{{Lorenz}, {Lunine}  \& {McKay}}{{Lorenz}
  et~al.}{1997}]{Lorenz1997}
{Lorenz} R.~D.,  {Lunine} J.~I.,   {McKay} C.~P.,  1997, \mn@doi [\grl]
  {10.1029/97GL52843}, \href
  {https://ui.adsabs.harvard.edu/abs/1997GeoRL..24.2905L} {24, 2905}

\bibitem[\protect\citeauthoryear{{Lovelock}}{{Lovelock}}{1965}]{Lovelock1965L}
{Lovelock} J.~E.,  1965, \mn@doi [\nat] {10.1038/207568a0}, \href
  {https://ui.adsabs.harvard.edu/abs/1965Natur.207..568L} {207, 568}

\bibitem[\protect\citeauthoryear{{Mennesson} et~al.,}{{Mennesson}
  et~al.}{2016}]{Mennesson2016}
{Mennesson} B.,  et~al., 2016, in \procspie. p. 99040L,
  \mn@doi{10.1117/12.2240457}

\bibitem[\protect\citeauthoryear{{Misra}, {Meadows}  \& {Crisp}}{{Misra}
  et~al.}{2014}]{Misra2014}
{Misra} A.,  {Meadows} V.,   {Crisp} D.,  2014, \mn@doi [\apj]
  {10.1088/0004-637X/792/1/61}, \href
  {https://ui.adsabs.harvard.edu/abs/2014ApJ...792...61M} {792, 61}

\bibitem[\protect\citeauthoryear{{O'Malley-James} \&
  {Kaltenegger}}{{O'Malley-James} \& {Kaltenegger}}{2018}]{OMalleyJames2018}
{O'Malley-James} J.~T.,  {Kaltenegger} L.,  2018, \mn@doi [Astrobiology]
  {10.1089/ast.2017.1798}, \href
  {https://ui.adsabs.harvard.edu/abs/2018AsBio..18.1123O} {18, 1123}

\bibitem[\protect\citeauthoryear{{Pavlov} \& {Kasting}}{{Pavlov} \&
  {Kasting}}{2002}]{Pavlov2002}
{Pavlov} A.~A.,  {Kasting} J.~F.,  2002, \mn@doi [Astrobiology]
  {10.1089/153110702753621321}, \href
  {https://ui.adsabs.harvard.edu/abs/2002AsBio...2...27P} {2, 27}

\bibitem[\protect\citeauthoryear{{Pavlov}, {Kasting}, {Brown}, {Rages}  \&
  {Freedman}}{{Pavlov} et~al.}{2000}]{Pavlov2000}
{Pavlov} A.~A.,  {Kasting} J.~F.,  {Brown} L.~L.,  {Rages} K.~A.,   {Freedman}
  R.,  2000, \mn@doi [\jgr] {10.1029/1999JE001134}, \href
  {https://ui.adsabs.harvard.edu/abs/2000JGR...10511981P} {105, 11981}

\bibitem[\protect\citeauthoryear{{Popp}, {Schmidt}  \& {Marotzke}}{{Popp}
  et~al.}{2016}]{Popp2016}
{Popp} M.,  {Schmidt} H.,   {Marotzke} J.,  2016, \mn@doi [Nature
  Communications] {10.1038/ncomms10627}, \href
  {https://ui.adsabs.harvard.edu/abs/2016NatCo...710627P} {7, 10627}

\bibitem[\protect\citeauthoryear{{Ramirez} \& {Kaltenegger}}{{Ramirez} \&
  {Kaltenegger}}{2016}]{Ramirez2016}
{Ramirez} R.~M.,  {Kaltenegger} L.,  2016, \mn@doi [\apj]
  {10.3847/0004-637X/823/1/6}, \href
  {https://ui.adsabs.harvard.edu/abs/2016ApJ...823....6R} {823, 6}

\bibitem[\protect\citeauthoryear{{Ramirez} \& {Kaltenegger}}{{Ramirez} \&
  {Kaltenegger}}{2017}]{Ramirez2017}
{Ramirez} R.~M.,  {Kaltenegger} L.,  2017, \mn@doi [\apjl]
  {10.3847/2041-8213/aa60c8}, \href
  {https://ui.adsabs.harvard.edu/abs/2017ApJ...837L...4R} {837, L4}

\bibitem[\protect\citeauthoryear{{Rodler} \& {L{\'o}pez-Morales}}{{Rodler} \&
  {L{\'o}pez-Morales}}{2014}]{Rodler2014}
{Rodler} F.,  {L{\'o}pez-Morales} M.,  2014, \mn@doi [\apj]
  {10.1088/0004-637X/781/1/54}, \href
  {https://ui.adsabs.harvard.edu/abs/2014ApJ...781...54R} {781, 54}

\bibitem[\protect\citeauthoryear{Rossow \& Schiffer}{Rossow \&
  Schiffer}{1999}]{RossowSchiffer1999}
Rossow W.~B.,  Schiffer R.~A.,  1999, Bull. Amer. Meteorol. Soc., 80, 2261

\bibitem[\protect\citeauthoryear{{Rugheimer} \& {Kaltenegger}}{{Rugheimer} \&
  {Kaltenegger}}{2018}]{Rugheimer2018}
{Rugheimer} S.,  {Kaltenegger} L.,  2018, \mn@doi [\apj]
  {10.3847/1538-4357/aaa47a}, \href
  {https://ui.adsabs.harvard.edu/abs/2018ApJ...854...19R} {854, 19}

\bibitem[\protect\citeauthoryear{{Rugheimer}, {Kaltenegger}, {Zsom}, {Segura}
  \& {Sasselov}}{{Rugheimer} et~al.}{2013}]{Rugheimer2013}
{Rugheimer} S.,  {Kaltenegger} L.,  {Zsom} A.,  {Segura} A.,   {Sasselov} D.,
  2013, \mn@doi [Astrobiology] {10.1089/ast.2012.0888}, \href
  {https://ui.adsabs.harvard.edu/abs/2013AsBio..13..251R} {13, 251}

\bibitem[\protect\citeauthoryear{{Rugheimer}, {Segura}, {Kaltenegger}  \&
  {Sasselov}}{{Rugheimer} et~al.}{2015a}]{Rugheimer2015Surface}
{Rugheimer} S.,  {Segura} A.,  {Kaltenegger} L.,   {Sasselov} D.,  2015a,
  \mn@doi [\apj] {10.1088/0004-637X/806/1/137}, \href
  {https://ui.adsabs.harvard.edu/abs/2015ApJ...806..137R} {806, 137}

\bibitem[\protect\citeauthoryear{{Rugheimer}, {Kaltenegger}, {Segura}, {Linsky}
   \& {Mohanty}}{{Rugheimer} et~al.}{2015b}]{Rugheimer2015Spectra}
{Rugheimer} S.,  {Kaltenegger} L.,  {Segura} A.,  {Linsky} J.,   {Mohanty} S.,
  2015b, \mn@doi [\apj] {10.1088/0004-637X/809/1/57}, \href
  {https://ui.adsabs.harvard.edu/abs/2015ApJ...809...57R} {809, 57}

\bibitem[\protect\citeauthoryear{{Schindler} \& {Kasting}}{{Schindler} \&
  {Kasting}}{2000}]{SchindlerKasting2000}
{Schindler} T.~L.,  {Kasting} J.~F.,  2000, \mn@doi [\icarus]
  {10.1006/icar.2000.6340}, \href
  {https://ui.adsabs.harvard.edu/abs/2000Icar..145..262S} {145, 262}

\bibitem[\protect\citeauthoryear{{Schwieterman}, {Cockell}  \&
  {Meadows}}{{Schwieterman} et~al.}{2015}]{Schwieterman2015}
{Schwieterman} E.~W.,  {Cockell} C.~S.,   {Meadows} V.~S.,  2015, \mn@doi
  [Astrobiology] {10.1089/ast.2014.1178}, \href
  {https://ui.adsabs.harvard.edu/abs/2015AsBio..15..341S} {15, 341}

\bibitem[\protect\citeauthoryear{{Segura}, {Krelove}, {Kasting}, {Sommerlatt},
  {Meadows}, {Crisp}, {Cohen}  \& {Mlawer}}{{Segura} et~al.}{2003}]{Segura2003}
{Segura} A.,  {Krelove} K.,  {Kasting} J.~F.,  {Sommerlatt} D.,  {Meadows} V.,
  {Crisp} D.,  {Cohen} M.,   {Mlawer} E.,  2003, \mn@doi [Astrobiology]
  {10.1089/153110703322736024}, \href
  {https://ui.adsabs.harvard.edu/abs/2003AsBio...3..689S} {3, 689}

\bibitem[\protect\citeauthoryear{{Segura}, {Kasting}, {Meadows}, {Cohen},
  {Scalo}, {Crisp}, {Butler}  \& {Tinetti}}{{Segura} et~al.}{2005}]{Segura2005}
{Segura} A.,  {Kasting} J.~F.,  {Meadows} V.,  {Cohen} M.,  {Scalo} J.,
  {Crisp} D.,  {Butler} R. A.~H.,   {Tinetti} G.,  2005, \mn@doi [Astrobiology]
  {10.1089/ast.2005.5.706}, \href
  {https://ui.adsabs.harvard.edu/abs/2005AsBio...5..706S} {5, 706}

\bibitem[\protect\citeauthoryear{{Segura}, {Meadows}, {Kasting}, {Crisp}  \&
  {Cohen}}{{Segura} et~al.}{2007}]{Segura2007}
{Segura} A.,  {Meadows} V.~S.,  {Kasting} J.~F.,  {Crisp} D.,   {Cohen} M.,
  2007, \mn@doi [\aap] {10.1051/0004-6361:20066663}, \href
  {https://ui.adsabs.harvard.edu/abs/2007A&A...472..665S} {472, 665}

\bibitem[\protect\citeauthoryear{{Segura}, {Walkowicz}, {Meadows}, {Kasting}
  \& {Hawley}}{{Segura} et~al.}{2010}]{Segura2010}
{Segura} A.,  {Walkowicz} L.~M.,  {Meadows} V.,  {Kasting} J.,   {Hawley} S.,
  2010, \mn@doi [Astrobiology] {10.1089/ast.2009.0376}, \href
  {https://ui.adsabs.harvard.edu/abs/2010AsBio..10..751S} {10, 751}

\bibitem[\protect\citeauthoryear{{Selsis}, {Kasting}, {Levrard}, {Paillet},
  {Ribas}  \& {Delfosse}}{{Selsis} et~al.}{2007}]{Selsis2007}
{Selsis} F.,  {Kasting} J.~F.,  {Levrard} B.,  {Paillet} J.,  {Ribas} I.,
  {Delfosse} X.,  2007, \mn@doi [\aap] {10.1051/0004-6361:20078091}, \href
  {https://ui.adsabs.harvard.edu/abs/2007A&A...476.1373S} {476, 1373}

\bibitem[\protect\citeauthoryear{{Shields} \& {Carns}}{{Shields} \&
  {Carns}}{2018}]{Shields2018}
{Shields} A.~L.,  {Carns} R.~C.,  2018, \mn@doi [\apj]
  {10.3847/1538-4357/aadcaa}, \href
  {https://ui.adsabs.harvard.edu/abs/2018ApJ...867...11S} {867, 11}

\bibitem[\protect\citeauthoryear{{Shields}, {Meadows}, {Bitz}, {Pierrehumbert},
  {Joshi}  \& {Robinson}}{{Shields} et~al.}{2013}]{Shields2013}
{Shields} A.~L.,  {Meadows} V.~S.,  {Bitz} C.~M.,  {Pierrehumbert} R.~T.,
  {Joshi} M.~M.,   {Robinson} T.~D.,  2013, \mn@doi [Astrobiology]
  {10.1089/ast.2012.0961}, \href
  {https://ui.adsabs.harvard.edu/abs/2013AsBio..13..715S} {13, 715}

\bibitem[\protect\citeauthoryear{{Shields}, {Bitz}, {Meadows}, {Joshi}  \&
  {Robinson}}{{Shields} et~al.}{2014}]{Shields2014}
{Shields} A.~L.,  {Bitz} C.~M.,  {Meadows} V.~S.,  {Joshi} M.~M.,   {Robinson}
  T.~D.,  2014, \mn@doi [\apjl] {10.1088/2041-8205/785/1/L9}, \href
  {https://ui.adsabs.harvard.edu/abs/2014ApJ...785L...9S} {785, L9}

\bibitem[\protect\citeauthoryear{{Snellen}, {de Kok}, {le Poole}, {Brogi}  \&
  {Birkby}}{{Snellen} et~al.}{2013}]{Snellen2013}
{Snellen} I.~A.~G.,  {de Kok} R.~J.,  {le Poole} R.,  {Brogi} M.,   {Birkby}
  J.,  2013, \mn@doi [\apj] {10.1088/0004-637X/764/2/182}, \href
  {https://ui.adsabs.harvard.edu/abs/2013ApJ...764..182S} {764, 182}

\bibitem[\protect\citeauthoryear{{Stevenson} et~al.,}{{Stevenson}
  et~al.}{2016}]{Stevenson2016}
{Stevenson} K.~B.,  et~al., 2016, \mn@doi [Publications of the Astronomical
  Society of the Pacific] {10.1088/1538-3873/128/967/094401}, \href
  {https://ui.adsabs.harvard.edu/abs/2016PASP..128i4401S} {128, 094401}

\bibitem[\protect\citeauthoryear{{Stubenrauch} et~al.,}{{Stubenrauch}
  et~al.}{2013}]{Stubenrauch2013}
{Stubenrauch} C.~J.,  et~al., 2013, \mn@doi [Bulletin of the American
  Meteorological Society] {10.1175/BAMS-D-12-00117.1}, \href
  {https://ui.adsabs.harvard.edu/abs/2013BAMS...94.1031S} {94, 1031}

\bibitem[\protect\citeauthoryear{{The LUVOIR Team}}{{The LUVOIR
  Team}}{2018}]{Luvoir2018}
{The LUVOIR Team} 2018, arXiv e-prints, \href
  {https://ui.adsabs.harvard.edu/abs/2018arXiv180909668T} {p. arXiv:1809.09668}

\bibitem[\protect\citeauthoryear{{Tinetti} et~al.,}{{Tinetti}
  et~al.}{2016}]{Tinetti2016}
{Tinetti} G.,  et~al., 2016, in \procspie. p. 99041X,
  \mn@doi{10.1117/12.2232370}

\bibitem[\protect\citeauthoryear{{Toon}, {McKay}, {Ackerman}  \&
  {Santhanam}}{{Toon} et~al.}{1989}]{Toon1989}
{Toon} O.~B.,  {McKay} C.~P.,  {Ackerman} T.~P.,   {Santhanam} K.,  1989,
  \mn@doi [\jgr] {10.1029/JD094iD13p16287}, \href
  {https://ui.adsabs.harvard.edu/abs/1989JGR....9416287T} {94, 16287}

\bibitem[\protect\citeauthoryear{{Traub} \& {Stier}}{{Traub} \&
  {Stier}}{1976}]{TraubStier1976}
{Traub} W.~A.,  {Stier} M.~T.,  1976, \mn@doi [\ao] {10.1364/AO.15.000364},
  \href {https://ui.adsabs.harvard.edu/abs/1976ApOpt..15..364T} {15, 364}

\bibitem[\protect\citeauthoryear{{Vladilo}, {Murante}, {Silva}, {Provenzale},
  {Ferri}  \& {Ragazzini}}{{Vladilo} et~al.}{2013}]{Vladilo2013}
{Vladilo} G.,  {Murante} G.,  {Silva} L.,  {Provenzale} A.,  {Ferri} G.,
  {Ragazzini} G.,  2013, \mn@doi [\apj] {10.1088/0004-637X/767/1/65}, \href
  {https://ui.adsabs.harvard.edu/abs/2013ApJ...767...65V} {767, 65}

\bibitem[\protect\citeauthoryear{{Williams} \& {Pollard}}{{Williams} \&
  {Pollard}}{2002}]{Williams2002}
{Williams} D.~M.,  {Pollard} D.,  2002, {Habitable Planets on Eccentric
  Orbits}.
p.~201

\bibitem[\protect\citeauthoryear{{Wolf} \& {Toon}}{{Wolf} \&
  {Toon}}{2015}]{Wolf2015}
{Wolf} E.~T.,  {Toon} O.~B.,  2015, \mn@doi [Journal of Geophysical Research
  (Atmospheres)] {10.1002/2015JD023302}, \href
  {https://ui.adsabs.harvard.edu/abs/2015JGRD..120.5775W} {120, 5775}

\bibitem[\protect\citeauthoryear{{Wordsworth} \& {Pierrehumbert}}{{Wordsworth}
  \& {Pierrehumbert}}{2013}]{Wordsworth2013}
{Wordsworth} R.~D.,  {Pierrehumbert} R.~T.,  2013, \mn@doi [\apj]
  {10.1088/0004-637X/778/2/154}, \href
  {https://ui.adsabs.harvard.edu/abs/2013ApJ...778..154W} {778, 154}

\bibitem[\protect\citeauthoryear{{Yang}, {Stancil}, {Balakrishnan}, {Forrey}
  \& {Bowman}}{{Yang} et~al.}{2013}]{Yang2013}
{Yang} B.,  {Stancil} P.~C.,  {Balakrishnan} N.,  {Forrey} R.~C.,   {Bowman}
  J.~M.,  2013, \mn@doi [\apj] {10.1088/0004-637X/771/1/49}, \href
  {https://ui.adsabs.harvard.edu/abs/2013ApJ...771...49Y} {771, 49}

\bibitem[\protect\citeauthoryear{Yang et~al.,}{Yang et~al.}{2016}]{Yang2016}
Yang J.,  et~al., 2016, \mn@doi [The Astrophysical Journal]
  {10.3847/0004-637x/826/2/222}, 826, 222

\bibitem[\protect\citeauthoryear{Zahnle, Claire  \& Catling}{Zahnle
  et~al.}{2006}]{Zahnle2006}
Zahnle K.,  Claire M.,   Catling D.,  2006, \mn@doi [Geobiology]
  {10.1111/j.1472-4669.2006.00085.x}, 4, 271

\bibitem[\protect\citeauthoryear{{Zsom}, {Kaltenegger}  \& {Goldblatt}}{{Zsom}
  et~al.}{2012}]{Zsom2012}
{Zsom} A.,  {Kaltenegger} L.,   {Goldblatt} C.,  2012, \mn@doi [\icarus]
  {10.1016/j.icarus.2012.08.028}, \href
  {https://ui.adsabs.harvard.edu/abs/2012Icar..221..603Z} {221, 603}

\makeatother
\end{thebibliography}







\bsp	
\label{lastpage}
\end{document}